\documentstyle[epsfig]{elsart}
\newcommand{\etmiss}{E_T\!\!\!\!\!\!\!\!\! \not \,\,\,\,\,\,\,}
\newcommand{\ptmiss}{p_T\!\!\!\!\!\!\!\!\! \not \,\,\,\,\,\,\,}
\newcommand{\mtrec}{m_t^{\mathrm rec}}
\newcommand{\mwrec}{M_W^{\mathrm rec}}
\newcommand{\mzrec}{M_Z^{\mathrm rec}}

\begin{document}
\begin{frontmatter}
\title{Multilepton production via top flavour-changing neutral
couplings at the CERN LHC}
\author{F. del Aguila \and J. A. Aguilar-Saavedra}
\address{Departamento de F\'{\i}sica Te\'{o}rica y del Cosmos \\
Universidad de Granada \\
E-18071 Granada, Spain}
\date{\today}
\begin{abstract}
$Zt$ and $\gamma t$ production with $Z \to l^+ l^-$ and $t \to Wb \to l \nu b$
provides the best determination of top flavour-changing neutral
couplings at the LHC. The bounds on $tc$ couplings eventually derived
from these processes
are similar to those expected from top decays, while the limits on $tu$
couplings
are better by a factor of two. The other significant $Z$ and $W$ decay modes
are also investigated.

\vspace*{0.5cm} \noindent
\end{abstract}
\end{frontmatter}
\section{Introduction}

Future colliders will explore higher energies looking for new physics. Even if
there is not a new production threshold the top quark will probe the physics
beyond the Standard Model (SM). Large $e^+ e^-$ and hadron colliders, in
particular the CERN Large Hadron Collider (LHC), will be top factories allowing
to measure the top couplings with high precision. Both types of machines are
complementary for if the LHC will produce many more tops, $e^+ e^-$ linear
colliders will face smaller backgrounds. Nevertheless the most realistic
estimates indicate that the precision which can be reached for flavour-changing
neutral (FCN) couplings can be up to a factor of three better at the LHC with a
luminosity of 100 fb$^{-1}$
\cite{papiro14,papiro15} than at
a 500 GeV $e^+ e^-$ linear collider with a luminosity of 100 fb$^{-1}$ (LC)
\cite{papiro16}. The top couplings which can signal to new physics can be
diagonal \cite{papiro1} or off-diagonal. We will concentrate
on the latter. FCN
couplings are very suppressed within the SM but can be large in simple
extensions \cite{papiro11}. Hence they are a good place to look for departures
from the SM. However present data only allow for FCN couplings large enough to
be observed at future colliders for the top quark.
In this paper we investigate the sensitivity of the LHC to
top FCN gauge couplings. This has already been studied for top decays 
\cite{papiro14,papiro15}. We look at $Vt$ production, $V=Z,\gamma$,
induced by
anomalous top couplings \cite{papiro2,papiro2b}, which, as we shall show,
provide better limits.

In order to describe FCN couplings between the top,
a light quark $q=u,c$ and a
$Z$ boson, a photon $A$ or a gluon $G^a$ we use the Lagrangian \cite{papiro10}
\begin{eqnarray}
-{\mathcal L} & = & \frac{g_W}{2 c_W} \bar t \gamma_\mu (X_{tq}^L P_L +
 X_{tq}^R P_R) q Z^\mu 
+ \frac{g_W}{2 c_W} \bar t (\kappa_{tq}^{(1)}-i \kappa_{tq}^{(2)} \gamma_5)
\frac{i \sigma_{\mu \nu} q^\nu}{m_t} q Z^\mu  \nonumber \\
& &  + e \bar t (\lambda_{tq}^{(1)}-i \lambda_{tq}^{(2)} \gamma_5)
\frac{i \sigma_{\mu \nu} q^\nu}{m_t} q A^\mu 
+ g_s \bar t (\zeta_{tq}^{(1)}-i \zeta_{tq}^{(2)} \gamma_5)
\frac{i \sigma_{\mu \nu} q^\nu}{m_t}
T^a q G^{a\mu}+{\mathrm h.c.}\,, \label{ec:1}
\end{eqnarray}
where $P_{R,L}=(1 \pm \gamma_5)/2$ and $T^a$ are the Gell-Mann matrices
satisfying ${\mathrm Tr}\, (T^a T^b) = \delta^{ab}/2$. The couplings are
constants corresponding to the first terms in the expansion in momenta.
This effective Lagrangian contains $\gamma_\mu$ terms of
dimension 4 and $\sigma_{\mu \nu}$ terms of dimension 5. The $\sigma_{\mu \nu}$
terms are the only ones allowed by the unbroken gauge symmetry, ${\mathrm
SU(3)}_c \times {\mathrm U(1)}_Q$. Due to their extra momentum factor they grow
with the energy and make large colliders the best place to measure them. They
are absent at tree-level in renormalizable theories like the SM, where they are
also suppressed by the GIM mechanism \cite{papiro4}. However the effective
couplings involving the top quark can be large in models with new physics near
the electroweak scale. In effective theories with only the SM light degrees of
freedom the $\gamma_\mu$ terms above also
result from dimension six operators after electroweak
symmetry breaking. However, if the scales involved are
similar, as happens with the top quark mass and the electroweak scale, the
$\gamma_\mu$ terms can be also large \cite{papiro3}.
This is the case in simple SM extensions with
vector-like quarks near the electroweak scale \cite{papiro6}. Although rare
processes strongly constrain FCN couplings between light quarks \cite{papiro5},
the top can have relatively large couplings to the quarks $u$ or $c$, 
but not to both simultaneously \cite{papiro11}.
Thus the third family seems the best place
to look for SM departures and Eq.~(\ref{ec:1}) is the lowest order contribution
to trilinear top FCN gauge couplings of any
possible extension. It is then important to measure them at the LHC
\cite{papiro30}.

\begin{figure}[htb]
\begin{center}
\mbox{\epsfig{file=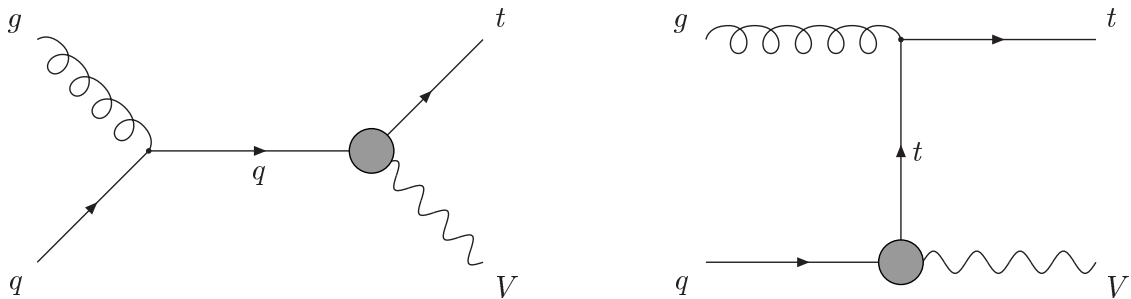,width=15cm}}
\end{center}
\caption{Feynman diagrams for $gq \to Vt$ via $Vtq$ FCN couplings
with $V=Z,\gamma$. The $Z$ boson and the top quark are off-shell and have the
SM decays.
\label{fig:feyn1} }
\end{figure}

\begin{figure}[htb]
\begin{center}
\mbox{\epsfig{file=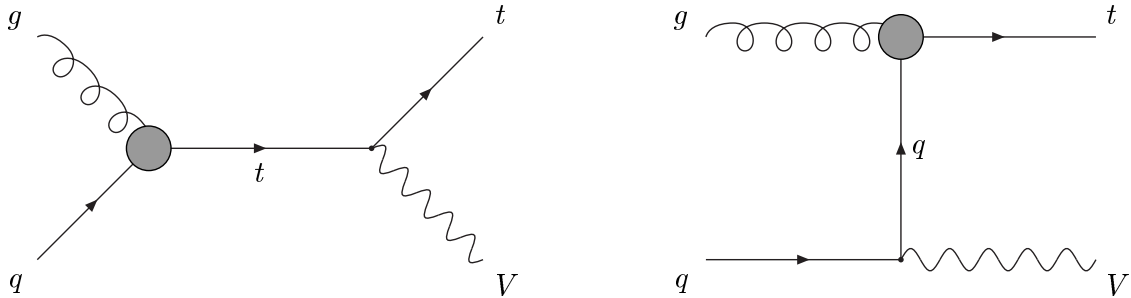,width=15cm}}
\end{center}
\caption{Feynman diagrams for $gq \to Vt$ via $gtq$ FCN couplings
with $V=Z,\gamma$. The $Z$ boson and the top quark are off-shell and have the
SM decays.
\label{fig:feyn2} }
\end{figure}

The vertices in Eq.~(\ref{ec:1}) are constrained by the nonobservation of the
top decays $t \to qV$
\cite{papiro12} and $t \to qg$ \cite{papiro13} at
Tevatron, implying the present direct limits
\begin{eqnarray}
X_{tq} & \equiv & \sqrt{|X_{tq}^L|^2+|X_{tq}^R|^2} \leq 0.84 \,, \nonumber \\
\kappa_{tq} & \equiv & \sqrt{|\kappa_{tq}^{(1)}|^2+|\kappa_{tq}^{(2)}|^2} 
\leq 0.78 \,, \nonumber \\
\lambda_{tq} & \equiv & \sqrt{|\lambda_{tq}^{(1)}|^2+|\lambda_{tq}^{(2)}|^2} 
\leq 0.33 \,, \nonumber \\
\zeta_{tq} & \equiv & \sqrt{|\zeta_{tq}^{(1)}|^2+|\zeta_{tq}^{(2)}|^2} 
\leq 0.15  \label{ec:2}
\end{eqnarray}
at 95\% C. L. (Unless otherwise stated, all bounds in this paper have this
confidence level.) The sensitivity of the LHC and LC for some of these
couplings has been studied for different processes in a series of papers.
At the LHC with an integrated
luminosity of 100 fb$^{-1}$ the
expected limits from top decays are
 $X_{tq} \leq 0.017$ \cite{papiro14}, $\lambda_{tq} \leq
0.0035$ \cite{papiro15}. The LC will probe the electroweak couplings in the
process $e^+ e^- \to t \bar c$, obtaining eventually
$X_{tq} \leq 0.051$, $\kappa_{tq} \leq 0.015$,
$\lambda_{tq} \leq 0.011$ \cite{papiro16}. The best limits on the $gtq$
vertices are expected from single top production
at LHC
\cite{papiro17}, $\zeta_{tu} \leq 0.0004$, $\zeta_{tc} \leq 0.0009$.
In the following we show that the process $gq \to Vt$ gives
competitive bounds on
the $Vtc$ couplings, $X_{tc} \leq 0.023$, $\kappa_{tc} \leq 0.016$, 
$\lambda_{tc} \leq 0.0065$, and the best limits on the $Vtu$ vertices,
$X_{tu} \leq 0.011$, $\kappa_{tu} \leq 0.0063$,  $\lambda_{tu} \leq 0.0021$. 
This process can take place through the s- and t-channel diagrams in
Fig.~\ref{fig:feyn1} via $Vtq$ couplings, or via $gtq$ couplings through
the diagrams in
Fig.~\ref{fig:feyn2}. However, in the second case the best limits on the strong
anomalous couplings
$\zeta_{tu} \leq 0.0018$, $\zeta_{tc} \leq 0.0037$ are 
less stringent than those derived from $tj$ production.

It must be noticed that all upper bounds we shall derive will be obtained
following the Feldman-Cousins construction \cite{papiro26} to obtain 95\%
C. L. intervals, assuming that the number of observed events $n$ equals the
expected background $n_b$. For the numerical evaluation of the confidence
intervals we use the PCI package \cite{papiro26b}. 
The intervals obtained with the Feldman-Cousins construction are similar to
standard 95\% C. L. upper limits on a Poisson variable with known background
\cite{papiro28} but they give less restrictive bounds
(see Table~\ref{tab:stat}).
Another possibility is to use the 95\% C. L. ($1.96 \,\sigma$) discovery
significance $n_s/\sqrt{n_b} \geq 1.96$, with $n_s$ the expected number of
signal events (see for instance Ref. \cite{papiro29}). The limits obtained are
similar to standard 95\% upper limits when the Poisson distribution can be
approximated by a Gaussian ($n_b \geq 5$). The quoted limits in Refs.
\cite{papiro14,papiro15,papiro16,papiro13,papiro17} have been obtained
using the statistical significance $n_s/\sqrt{n_s+n_b} \geq 3$.
This is a conservative estimate with a 99\% C. L. and weakens the
bounds on the anomalous couplings given here by a
factor $\sim \sqrt 2$ (see Table~\ref{tab:stat}). This is part of the
improvement found.

\begin{table}[htb]
\begin{center}
\begin{tabular}{ccccc}
$n_b$ & $n_s$ (Ref. \cite{papiro26}) & $n_s$ (Ref. \cite{papiro28}) &
$n_s$ (Ref. \cite{papiro29}) & 
$n_s$ (Refs. \cite{papiro14,papiro15,papiro16,papiro13,papiro17}) \\
0 & 3.09 & 3.00 & 0 & 3.84 \\
5 & 6.26 & 5.51 & 4.38 & 6.71 \\
10 & 7.82 & 6.97 & 6.20 & 8.41 \\
15 & 9.31 & 8.09 & 7.59 & 9.75 
\end{tabular}
\end{center}
\caption{95\% C. L. upper bounds on signal events $n_s$ for samples of $n_b$ 
backgound events obtained 
using four different prescriptions.}
\label{tab:stat}
\end{table}

The sensitivity to top FCN couplings varies with the $Z$ and $W$ decay
modes. These are collected in Table~\ref{tab:chan},
together with their branching fractions and main backgrounds. We neglect
nonstandard top decays since they are a small fraction to start with. Moreover,
this approximation becomes better when smaller are the upper limits on anomalous
decays, which is the practical case we are interested in. In the
leptonic modes
we only consider decays to electrons and muons, but a good $\tau$ tagging
efficiency increases the leptonic branching ratios in Table~\ref{tab:chan} and
improves their statistical significance. The relevance of the different decay
channels results from the balance between the size of the signals, the
corresponding backgrounds and their statistical significance, and varies
substantially from
the Fermilab Tevatron to the LHC. For instance, at
Tevatron Run II with a luminosity of 2 fb$^{-1}$ the
$\nu \bar \nu jjb$ decay mode is the most
interesting one due to its branching ratio, with 235 signal and 22 background
events after kinematical cuts for $X_{tu}=0.84$ \cite{papiro2}. At this
collider, the $l^+ l^- l \nu b$
channel has a much larger signal to background ratio (28 signal and 0.07
background events {\em before} kinematical cuts for $X_{tu}=0.84$) but still its
statistical significance is lower and the bound obtained from the former decay
mode is more restrictive. However, this behaviour is reversed at the LHC, where
the increase in energy and luminosity
makes the $l^+ l^- l \nu b$ channel the most sensitive one.

Other decay modes also provide stringent limits. The $l^+ l^- jjb$ channel gives
competitive but worse results than $l^+ l^- l \nu b$ at the LHC.
The modes with hadronic $Z$ decay, in particular
$jj l \nu b$, which gives significant bounds at Tevatron, have huge
backgrounds at higher energies and luminosities:
the $Wt$ background becomes more important due to the larger
$b$ content of the proton
at LHC energies, and $t \bar t$, with $t(\bar t) \to jjb$,
$\bar t (t) \to l \nu b$
and one jet undetected, also grows rapidly with the energy.
Thus the $jj l\nu b$ channel is not interesting any more.
The situation improves considering only $Z \to b \bar b$ and requiring three $b$
tags, but the results are worse than for the channels with
$Z \to l^+ l^-$. The $\nu \bar \nu jjb$ mode also has as backgrounds $Wt$ and
$t \bar t$ but we will discuss this signal for
illustration, since this mode is the best one at Tevatron. The
remaining decay channels
$\nu \bar \nu l \nu b$, $jjjjb$ and $b \bar b jj b$
have too large backgrounds and will not be treated here. The $\gamma l \nu b$
decay mode
is the most interesting one in $\gamma t$ production, due to its small
$\gamma W j$ background. $\gamma jjb$ gives also significant but less
restrictive limits.

It is worth to note that $b$ tagging plays an essential r\^{o}le in enhancing
single top signals. The approximately conserved $b$-Parity in the SM
\cite{papiro7} allows in practice to get rid of large backgrounds with an even
number of $b$ quarks in the final state.

\begin{table}[htb]
\begin{center}
\begin{tabular}{lllcllc}
& \multicolumn{3}{c}{$W \to l \nu$} & 
\multicolumn{3}{c}{$W \to jj$} \\
& \multicolumn{2}{c}{Final state} & Backgrounds
& \multicolumn{2}{c}{Final state} & Backgrounds \\
$Z \to l^+ l^-$ & 
$l^+ l^- l \nu b$    & $1.5\%$ & $ZWj$ &
$l^+ l^- jj b$  & $4.6\%$ & $Zjjj$, $ZWj$ \\

$Z \to \nu \bar \nu$ &
$ \nu \bar \nu l \nu b$ & $4.3\%$  & $Wj$ &
$ \nu \bar \nu jj b$  & $13.6\%$ & $t \bar t$,
$W t$, $Zjjj$ \\

$Z \to b \bar b$ &
$b \bar b l \nu b$ & $3.3\%$ & $t \bar t$, $W t$, $ZWj$, $Wb \bar b j$&
$b \bar b jj b$ & $10.2\%$ & $b \bar b jjj$ \\

$Z \to jj$ &
$jj l \nu b$ & $15.2\%$ & $t \bar t$, $Wt$, $Wjjj$ &
$jjjjb$ & $47.4\%$ & $jjjjj$ \\
\hline
$\gamma$ (stable) &
$\gamma l \nu b$ & $21.8\%$ & $\gamma W j$ &
$\gamma jj b$ & $67.8\%$ & $\gamma jjj$, $\gamma Wj$
\end{tabular}
\caption{Branching ratios and main backgrounds for the different $Z$ and $W$
decay channels. $j$ and $jj$ include $b$ and $b \bar b$ when possible (we
neglect
the small Cabibbo-Kobayashi-Maskawa mixing angles $V_{ub}, V_{cb} \sim 0$
\cite{papiro8}).
\label{tab:chan}}
\end{center}
\end{table}

This paper is organized as follows. In Section 2 we discuss how the different
signals and backgrounds are generated. In Sections 3 and 4 we analyze $Zt$
and $\gamma t$ signals. We summarize in Section 5.

\section{Signal and background simulation}

In general $Zt$ production yields five fermion final states with
at least one $b$ quark $f\! f\! f\! fb$.
We evaluate these signals with the exact matrix element for the s- and
t-channel diagrams $gq \to Zt \to ZWb \to f\! f\! f\! fb$
(see Figs.~\ref{fig:feyn1} and
\ref{fig:feyn2}). The SM diagrams $gq \to ZWb$
are much smaller in the phase space region
of interest and suppressed by small mixing angles, so we
neglect them in the signal evaluation. We also ignore interferences from
identical fermion interchange in
the final state. All fermions are assumed massless except the top quark, and we
assume only one type of coupling to be nonzero at a time. We
evaluate $\gamma t$ production in a similar way, with the s- and t-channel
diagrams $gq \to \gamma t \to \gamma f\! fb$. $t$ and $\bar t$ production are
summed up in all cases.

The $VWj$ backgrounds are evaluated considering $g q_u \to V W q_d$,
$g q_d \to V W q_u$ and $q_u \bar q_d \to V W g$ (with $q_u=u,c$ and
$q_d=d,s,b$) plus the charge conjugate processes. We calculate the matrix
elements for $g q_u \to V W q_d$, including the eight SM diagrams and decaying
the $Z$ and $W$ afterwards. The matrix elements for the two other processes are
obtained by crossing symmetry. Our results
are consistent with those given in Ref. \cite{papiro18}.

In order to calculate the $Zjjj$, $Zb \bar bj$, $Wjjj$ and $Wb \bar bj$
backgrounds we use VECBOS \cite{papiro19} modified to include the energy
smearing and trigger and kinematical cuts. We also include routines to generate
the kinematical distributions. To generate the $\gamma jjj$ background we have
further modified VECBOS to produce photons instead of $Z$ bosons. 
This is done by introducing a `photon' with a small mass
$m_\gamma=0.1$ GeV and substituting the $Z$ couplings by the photon couplings
everywhere. The total width of such `photon' is calculated to be
$\Gamma_\gamma = 1.73 \cdot 10^{-3}$ GeV, with an $e^+ e^-$ branching ratio
equal to 0.15. We have checked that the results are the same for a heavier
`photon' with $m_\gamma=1$ GeV and $\Gamma_\gamma = 1.73 \cdot 10^{-2}$ GeV.

We have also to evaluate $Wt$ production, which is similar to the signal $Zt$,
but proceeds through the SM process $gb \to Wt \to W W b \to f\! f\! f\! fb$,
and is calculated analogously. $t \bar t$
production with some particles missed by the detector must be also considered
\cite{papiro29b}.
Finally, $Wj$,
$jjjjj$ and $b \bar b jjj$ production is so large that makes unnecessary a
detailed discussion of the corresponding signals and of the backgrounds
themselves. Other possible small backgrounds are neglected
\cite{papiro2b,papiro29}.

We include throughout the paper a $K$ factor equal to 1.1 for all processes
\cite{papiro20} except for $t \bar t$ production for which we assume $K=2.0$
\cite{papiro21}. We use MRST structure functions set A \cite{papiro22} with
$Q^2 = \hat s$. The cross sections have some dependence on the structure
functions chosen but not the trend of our results.

After generating
signals and backgrounds we imitate the experimental conditions with a Gaussian
smearing of the lepton ($l$), photon ($\gamma$) and jet ($j$) energies
\cite{papiro31},
\begin{equation}
\frac{\Delta E^{l,\gamma}}{E^{l,\gamma}} = \frac{10\%}{\sqrt{E^{l,\gamma}}}
\oplus 0.3\% \,, ~~~~
\frac{\Delta E^{j}}{E^{j}} = \frac{50\%}{\sqrt{E^j}} \oplus 3\% \,,
\end{equation}
where the energies are in GeV and
the two terms are added in quadrature. For simplicity we assume that the
energy smearing for muons is the same as for electrons. We then apply
detector cuts on transverse momenta $p_T$, pseudorapidities $\eta$ and distances
in $(\eta,\phi)$ space $\Delta R$:
$$
p_T^l \geq 15 {\mathrm ~ GeV} ~,~~ p_T^j \geq 20 {\mathrm ~ GeV} ~,~~ 
p_T^\gamma \geq 40 {\mathrm ~ GeV}
$$
\begin{equation}
|\eta^{l,j,\gamma}|  \leq 2.5 ~,~~
\Delta R_{jj,lj,\gamma l,\gamma j}  \geq 0.4 \,.
\end{equation}
For the $Wt$ and $t \bar t$ backgrounds, we estimate in how many events we
miss the
charged lepton and the $b$ or any other jet demanding that their momenta
and pseudorapidities
satisfy $p_T^l < 15$ GeV, $p_T^j < 20$ GeV or $|\eta^{l,j}| > 3$.

For the events to be triggered, we require both the signal and background
to fulfil at least one of the trigger conditions \cite{papiro23}. For the first
LHC Run with a luminosity of 10 fb$^{-1}$ (L),
\begin{itemize}
\item one jet with $p_T \geq 180$ GeV,
\item three jets with $p_T \geq 75$ GeV,
\item one charged lepton with $p_T \geq 20$ GeV,
\item two charged leptons with $p_T \geq 15$ GeV,
\item one photon with $p_T \geq 40$ GeV,
\item missing energy $\etmiss \geq 50$ GeV and one jet with $p_T \geq 50$ GeV,
\end{itemize}
and for the second Run with 100 fb$^{-1}$ (H),
\begin{itemize}
\item one jet with $p_T \geq 290$ GeV,
\item three jets with $p_T \geq 130$ GeV,
\item one charged lepton with $p_T \geq 30$ GeV,
\item two charged leptons with $p_T \geq 20$ GeV,
\item one photon with $p_T \geq 60$ GeV,
\item missing energy $\etmiss \geq 100$ GeV and one jet with $p_T \geq 100$ GeV.
\end{itemize}

Finally, we require a tagged $b$
jet in the final state taking advantage
of a good $b$ tagging efficiency
$\sim 60\%$ perhaps better than the one finally achieved \cite{papiro24}.
There is also a small probability $\sim 1\%$ that a
jet which does not result from the fragmentation of a $b$ quark
is misidentified as a $b$ jet \cite{papiro25}.
$b$ tagging is then implemented in the
Monte Carlo routines taking into account all possibilities of $b$
(mis)identification. As we shall see, this reduces substantially the
backgrounds.

\begin{figure}[htb]
\begin{center}
\mbox{
\epsfig{file=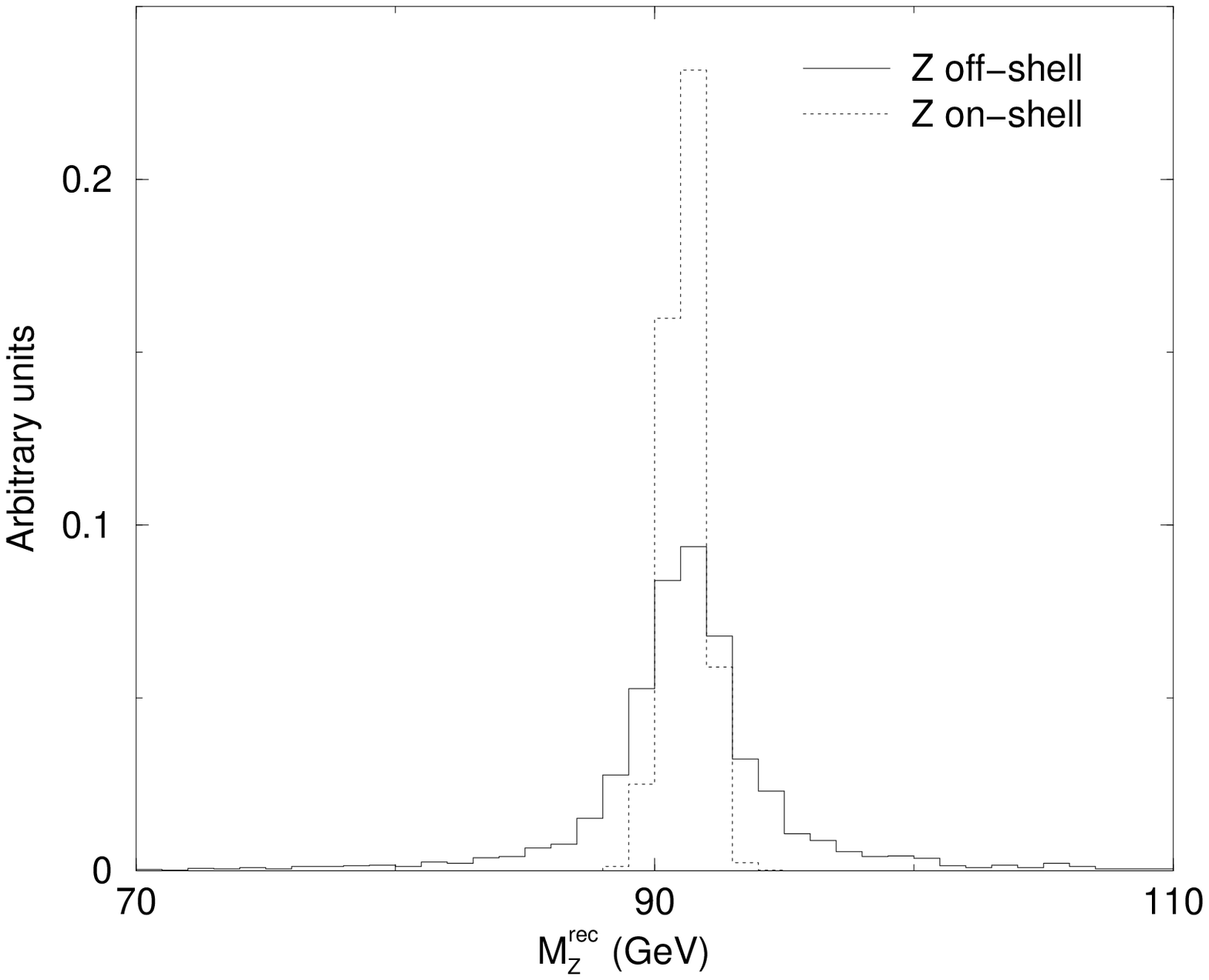,width=10cm}}
\caption{Reconstructed $Z$ mass $\mzrec$ distribution for the $l^+ l^- jjb$
signal at LHC with Gaussian energy smearing and detector and trigger cuts.
\label{fig:zon}}
\end{center}
\end{figure}

To conclude this Section let us emphasize the importance of performing the
full $2 \to n$
body calculation with the intermediate particles off-shell.
To illustrate
the relative importance of considering the
intermediate particles off-shell and
of simulating the detector conditions
with a Gaussian smearing of jet, charged lepton and photon energies we
consider the reconstructed $Z$ boson mass $\mzrec$ for its
leptonic decay, with $\mzrec$ defined as the two lepton invariant mass (for
instance in the $Zt$ decay mode $l^+ l^- jjb$).
In Fig.~\ref{fig:zon} we plot the corresponding
distributions for $Z$ off- and on-shell, including in both cases the energy
smearing. We observe that at LHC, for these detector
resolutions and lepton energies, the
effect of the off-shellness is more important than the energy smearing. 
(Of course the same applies to the intermediate $t$ and $W$.)
Applying kinematical cuts on reconstructed masses (or any related
variable such as the sum of $p_T$'s of the decay products) without allowing
the corresponding particles to be off-shell at least in the signal
would lead to very optimistic limits. In the case
of the $Z$ hadronic decay, the only way to distinguish
$Zt$ production from the copious $Wt$
production is requiring a $Z$ reconstructed
mass not consistent with the $W$ mass. Hence it is essential to generate
this signal and background with $Z$ and $W$ off-shell.

\section{$Zt$ production}

Although anomalous $Zt$ production is a tree level process with a strong vertex
(see Fig.~\ref{fig:feyn1}) we will be eventually interested in small anomalous
$Ztq$ couplings which give small cross sections.
It is then important to perform a detailed analysis
to look for the statistically most
significant decay channels. As we will show, these are the modes with
$Z$ decaying leptonically, $l^+ l^- l \nu b$ and $l^+ l^- jjb$.
We discuss the
two channels in turn. Of the modes with $Z$ decaying hadronically,
the $jj l \nu b$ decay mode (which has a statistical significance similar
to $l^+ l^- l \nu b$ at Tevatron Run II) has in practice too
large $Wt$ and $t \bar t$ backgrounds at the LHC to be interesting. However,
considering only $Z \to b \bar b$ and then requiring three $b$ tags
these backgrounds are reduced and the $b \bar b l \nu b$ signal becomes the most
relevant one with $Z$ decaying into hadrons. Finally we discuss the
$\nu \bar \nu jjb$ signal, which is the best one at Tevatron Runs I and II but
at LHC has too large $Wt$ and $t \bar t$ backgrounds. For each decay channel
limits on anomalous $gtq$ couplings can be derived through the process in
Fig.~\ref{fig:feyn2}. We quote without discussion the best limits from $Zt$
production, which are also provided by the three charged lepton decay
$l^+ l^- l \nu b$.

\subsection{$l^+ l^- l \nu b$ signal}
\begin{figure}[htb]
\begin{center}
\mbox{
\epsfig{file=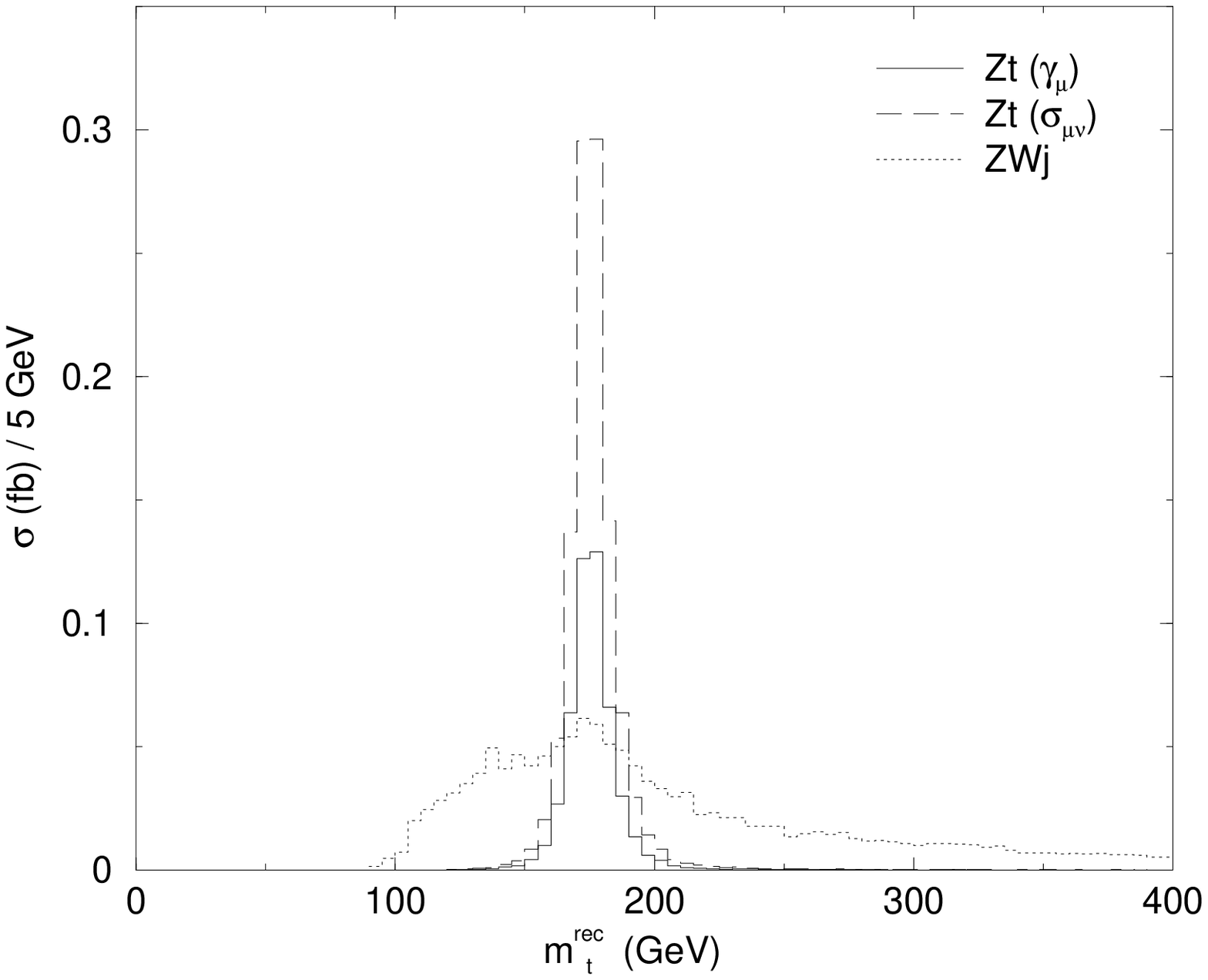,width=10cm}}
\caption{Reconstructed top mass $\mtrec$ distribution before kinematical cuts
for the $gu \to l^+ l^- l \nu b$ signal and background in LHC Run L.
We use  $X_{tu} = 0.02$, $\kappa_{tu}=0.02$ \label{fig:eee-mt}}
\end{center}
\end{figure}

\begin{figure}[htb]
\begin{center}
\mbox{
\epsfig{file=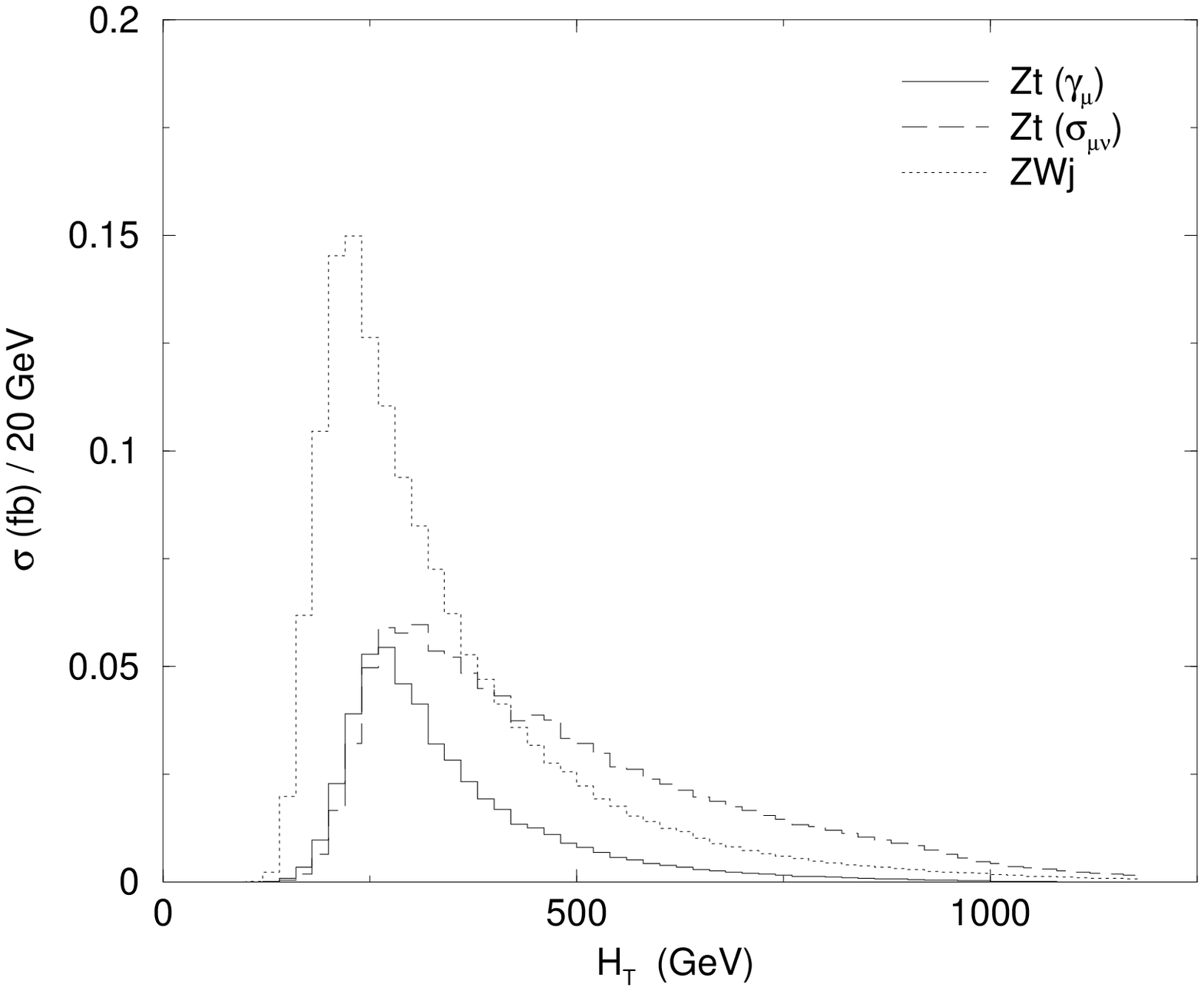,width=10cm}}
\caption{Total transverse energy $H_T$ distribution before kinematical cuts
for the $gu \to l^+ l^- l \nu b$ signal and background in LHC Run L.
We use  $X_{tu} = 0.02$, $\kappa_{tu}=0.02$ \label{fig:eee-ht}}
\end{center}
\end{figure}

\begin{figure}[htb]
\begin{center}
\mbox{
\epsfig{file=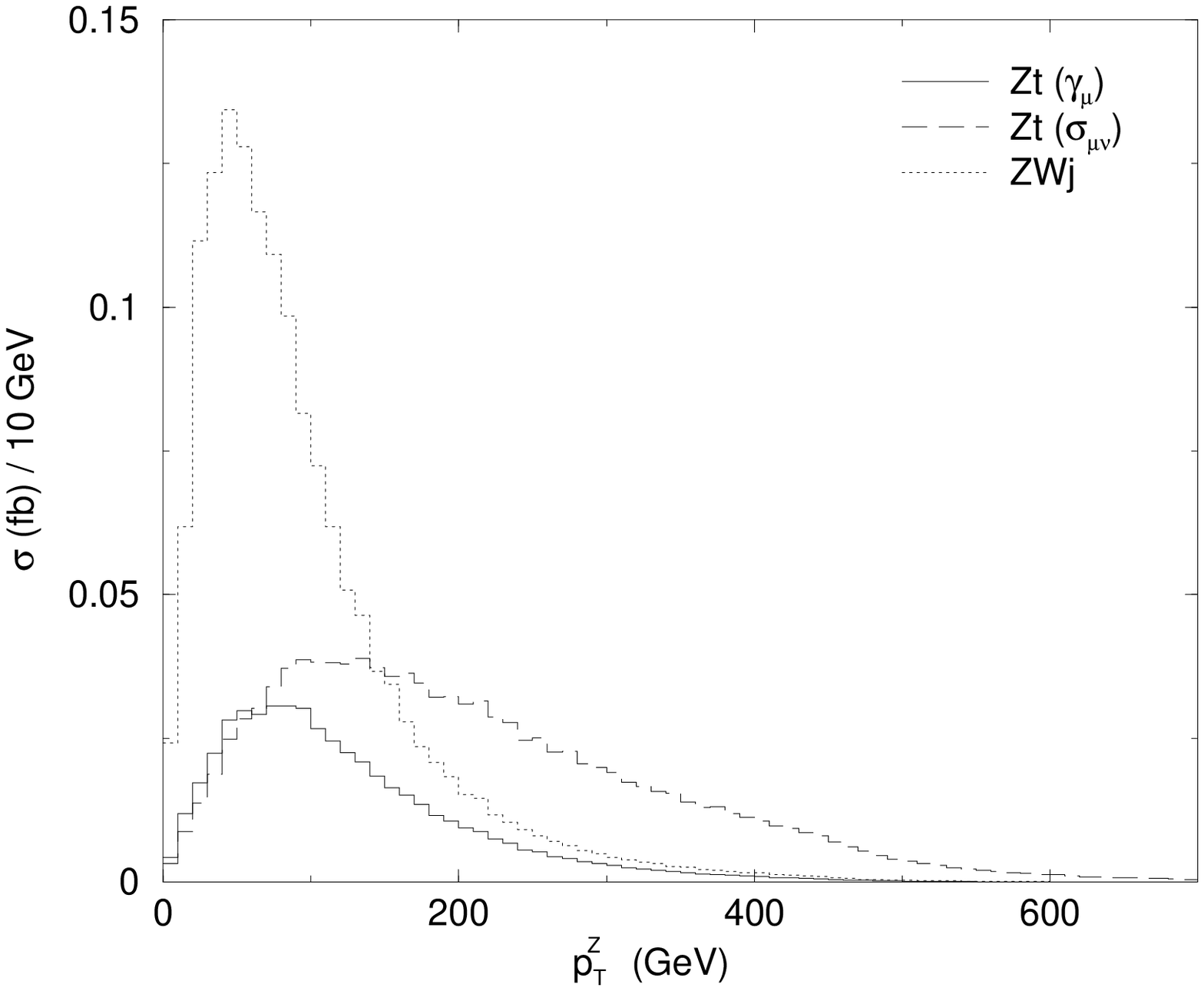,width=10cm}}
\caption{$p_T^Z$ distribution before kinematical cuts
for the $gu \to l^+ l^- l \nu b$ signal and background in LHC Run L.
We use  $X_{tu} = 0.02$, $\kappa_{tu}=0.02$ \label{fig:eee-ptz}}
\end{center}
\end{figure}

The $l^+ l^- l \nu b$ mode is the best decay channel to search for anomalous
$Ztq$ couplings at LHC. Its branching ratio is very small, $1.5\%$, but its
only background is $ZWj$ production, with
$j$ misidentified as a $b$ with a probability of 0.01.
The true $b$ production from initial $u$ and $c$
quarks is suppressed by the Cabibbo-Kobayashi-Maskawa matrix elements
$|V_{ub}|^2$ and $|V_{cb}|^2$ \cite{papiro8},
respectively, and is negligible. For better
comparison here and throughout this paper 
we normalize the signal to $X_{tq} = 0.02$ and $\kappa_{tq}=0.02$, and
to be definite we fix the ratio $X_{tq}^L / X_{tq}^R = 4/3$.
To perform kinematical cuts on the signal and background we must first identify
the pair of oppositely charged leptons $l^+ l^-$
 resulting from the $Z$ decay.
There are two such pairs and we take that with invariant mass $\mzrec$
closest to the $Z$ mass. We do not perform any kinematical cut on $\mzrec$
because obviously signal and background peak around $M_Z$ and use this
procedure only to identify the charged lepton $l$
resulting from the $W$ decay.
We then make the hypothesis that all missing energy comes from a single
neutrino with $p^\nu=(E^\nu,\ptmiss,p_L^\nu)$, and $\ptmiss$ the missing
transverse momentum. Using $(p^l + p^\nu)^2 = M_W^2$ we find two solutions for
$p^\nu$, and we choose the one making the reconstructed top mass
$\mtrec \equiv \sqrt{(p^l +  p^\nu + p^b)^2}$ closest to $m_t$. In
Fig.~\ref{fig:eee-mt} we plot this distribution for the $gu \to Zt$
signal and background. We observe that the background has a maximum near
$m_t$. This is because in this $m_t$ reconstruction method we first impose
$(p^l + p^\nu)^2 = M_W^2$ and then we choose the best of the two possible
$\mtrec$ values.
Other interesting kinematical variables are the total transverse energy $H_T$ in
Fig.~\ref{fig:eee-ht}, defined in general
as the scalar sum of the $p_T$'s of all jets, photons and charged leptons
plus $\etmiss$, and $p_T^Z$, the reconstructed transverse momentum
of the $Z$ boson,
plotted in Fig.~\ref{fig:eee-ptz}.

To enhance the signal to background ratio we apply the kinematical cuts on
$\mtrec$, $H_T$ and $p_T^Z$ in Table~\ref{tab:eee1}. In addition we require
$\etmiss > 5$ GeV to ensure a meaningful top mass reconstruction.
The higher luminosity of
Run H allows more stringent cuts that eliminate $90\%$ of the background while
retaining more than $60\%$ of the signal. The total number of signal and
background events for Runs L and H with integrated luminosities of 10 fb$^{-1}$
and 100 fb$^{-1}$, respectively, is collected in Table~\ref{tab:eee2}, using
for the signal $X_{tq}=0.02$, $\kappa_{tq}=0.02$. Note that
for Run L the trigger is redundant since all events passing the detector cut
$p_t^l \geq 15$ GeV automatically satisfy the leptonic trigger.
Comparing the numbers before kinematical cuts it can be also observed that
in Run H the trigger has little effect, due to the presence of three
charged leptons in the final state.
To derive upper bounds on the coupling
constants we use the prescriptions of Ref. \cite{papiro26} (similar to those
applied in Ref. \cite{papiro12} to obtain the present Tevatron limits). The
contributions from $u$ and $c$ quarks
must be summed up if a positive signal is
observed. However, if there is no evidence for this process, independent
bounds for each quark and coupling can be obtained,
$X_{tu} \leq 0.022$, $X_{tc} \leq 0.045$,
$\kappa_{tu} \leq 0.014$, $\kappa_{tc} \leq 0.034$ after Run L and
$X_{tu} \leq 0.011$, $X_{tc} \leq 0.023$,
$\kappa_{tu} \leq 0.0063$, $\kappa_{tc} \leq 0.016$ after Run H. (The expected
limit from top decay after Run H is $X_{tq} \leq 0.017$.)

\begin{table}[htb]
\begin{center}
\begin{tabular}{ccc}
Variable & Run L & Run H \\
$\mtrec$ & 150--200 & 160--190 \\
$H_T$ & $>200$ & $>260$ \\
$p_T^Z$ & & $>50$ \\
\end{tabular}
\caption{Kinematical cuts for the $l^+ l^- l \nu b$ decay channel. The masses,
energies and momentum are in GeV. \label{tab:eee1}}
\end{center}
\end{table}

\begin{table}[htb]
\begin{center}
\begin{tabular}{ccccc}
 & \multicolumn{2}{c}{Run L} & \multicolumn{2}{c}{Run H} \\
& before & after & before & after \\[-0.4cm]
& cuts & cuts & cuts & cuts \\
$gu \to Zt\, (\gamma_\mu)$ & 5.0 & 4.8 & 49.4 & 31.6 \\
$gc \to Zt\, (\gamma_\mu)$ & 1.1 & 1.1 & 11.4 & 6.5 \\
$gu \to Zt\, (\sigma_{\mu \nu})$ & 11.1 & 10.9 & 111 & 88.1 \\
$gc \to Zt\, (\sigma_{\mu \nu})$ & 2.0 & 1.9 & 19.5 & 14.4 \\
$ZWq_u$ & 4.9 & 1.4 & 49.2 & 5.0 \\
$ZWq_d$ & 5.5 & 1.4 & 54.8 & 5.1 \\
$ZWg$ & 4.7 & 1.1 & 47.4 & 4.0
\end{tabular} 
\caption {Number of $l^+ l^- l \nu b$ events before and after
the kinematical cuts in Table~\ref{tab:eee1} for the $Zt$ signal and
backgrounds. We use $X_{tq}=0.02$ and $\kappa_{tq}=0.02$.  \label{tab:eee2}}
\end{center}
\end{table}

One may wonder whether it would be useful to exploit the characteristic $q^\nu$
behaviour of the $\sigma_{\mu \nu}$ couplings requiring large transverse momenta
to obtain more stringent bounds. In this case, it makes little difference and
requiring $H_T > 360$ GeV in Run H only reduces the $\kappa_{tu}$ limit to
0.006.

This decay channel can be also used to constrain the $gtq$ couplings
 through the
process in Fig.~\ref{fig:feyn2}. Proceeding in the same way as before we obtain
$\zeta_{tu} \leq 0.0069 (0.0030)$, $\zeta_{tc} \leq 0.017 (0.0078)$ after
Run L (H).

\subsection{$l^+ l^- jjb$ signal}

This is the most interesting channel  with hadronic $W$ decay. At
Tevatron this mode is surpassed by the $\nu \bar \nu jjb$ channel due to the
relatively low statistics available and its greater branching ratio,
but this is not the case at LHC. The main
background for $l^+ l^- jjb$ is $Zjjj$ production
with a jet misidentified as a $b$. The
second background is $Z b \bar b j$ production with only one $b$ tagged.
The $ZWj$ background in this case is much smaller but we take it into account at
the end for comparison.

\begin{figure}[htb]
\begin{center}
\mbox{
\epsfig{file=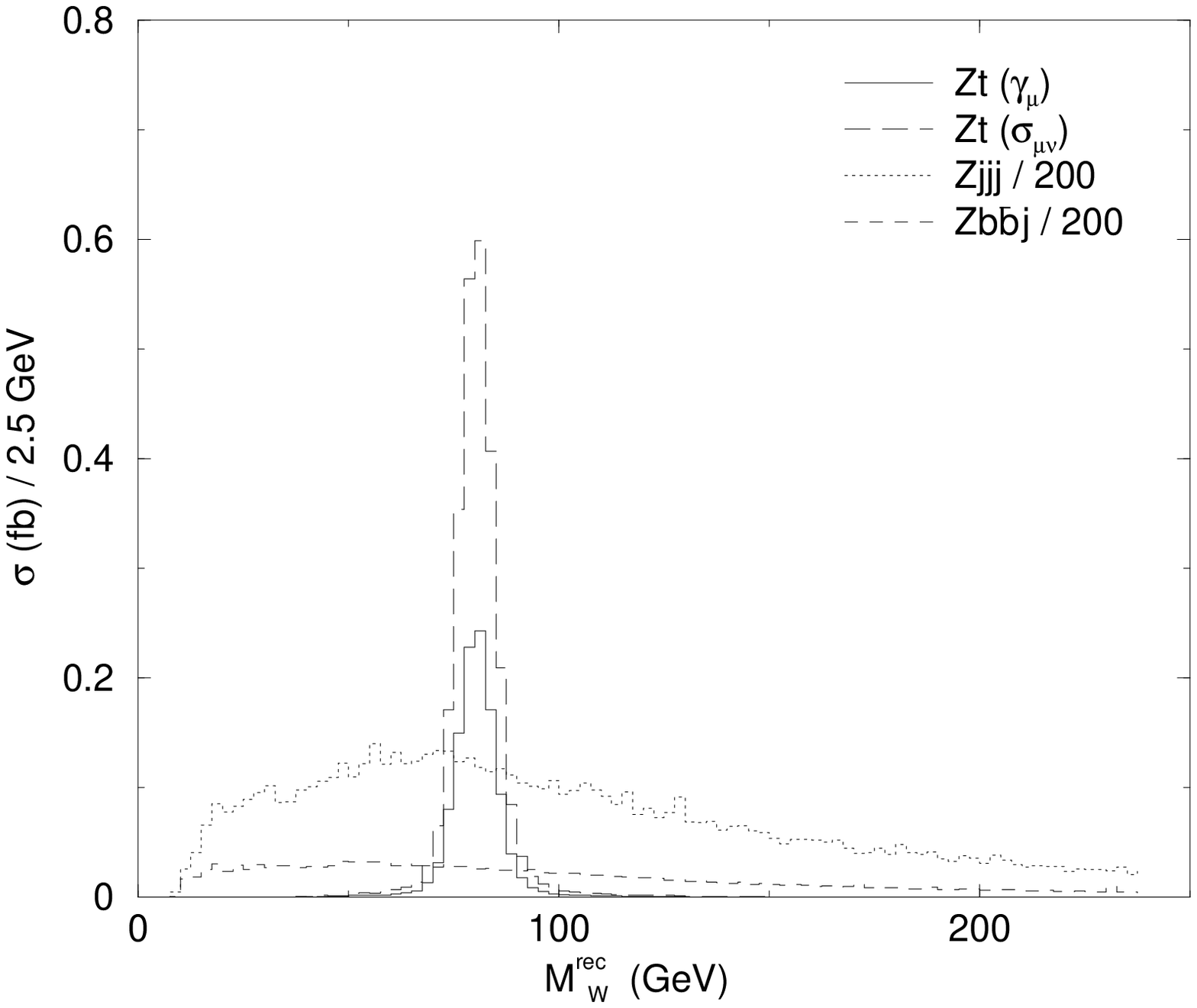,width=10cm}}
\caption{Reconstructed $W$ mass $\mwrec$ distribution before kinematical cuts
for the $gu \to l^+ l^- jjb$ signal and backgrounds in LHC Run L. We use 
$X_{tu} = 0.02$, $\kappa_{tu}=0.02$ \label{fig:eejj-mw}}
\end{center}
\end{figure}

\begin{figure}[htb]
\begin{center}
\mbox{
\epsfig{file=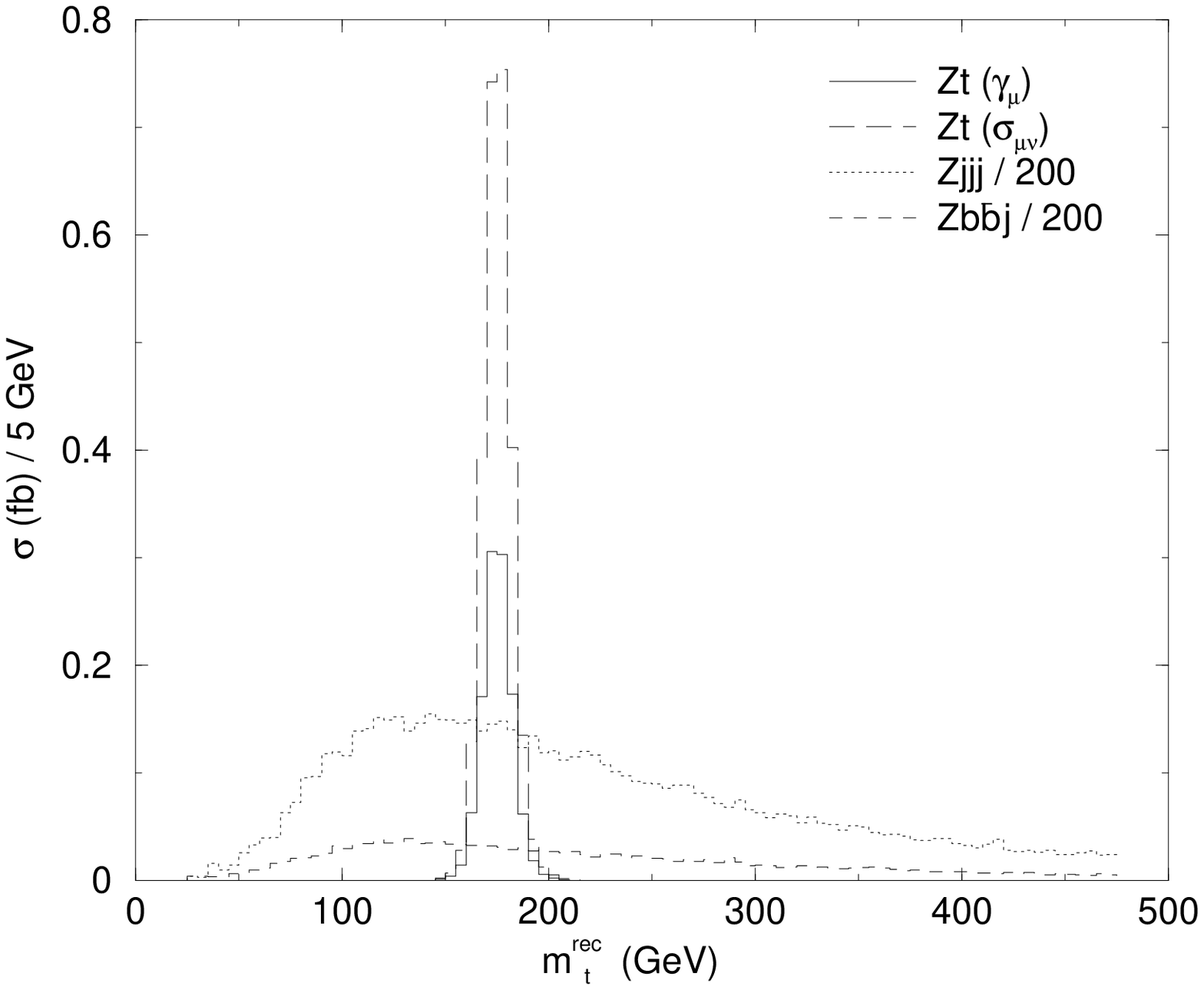,width=10cm}}
\caption{Reconstructed top mass $\mtrec$ distribution before kinematical cuts
for the $gu \to l^+ l^- jjb$ signal and backgrounds in LHC Run L. We use 
$X_{tu} = 0.02$, $\kappa_{tu}=0.02$ \label{fig:eejj-mt}}
\end{center}
\end{figure}

\begin{figure}[htb]
\begin{center}
\mbox{
\epsfig{file=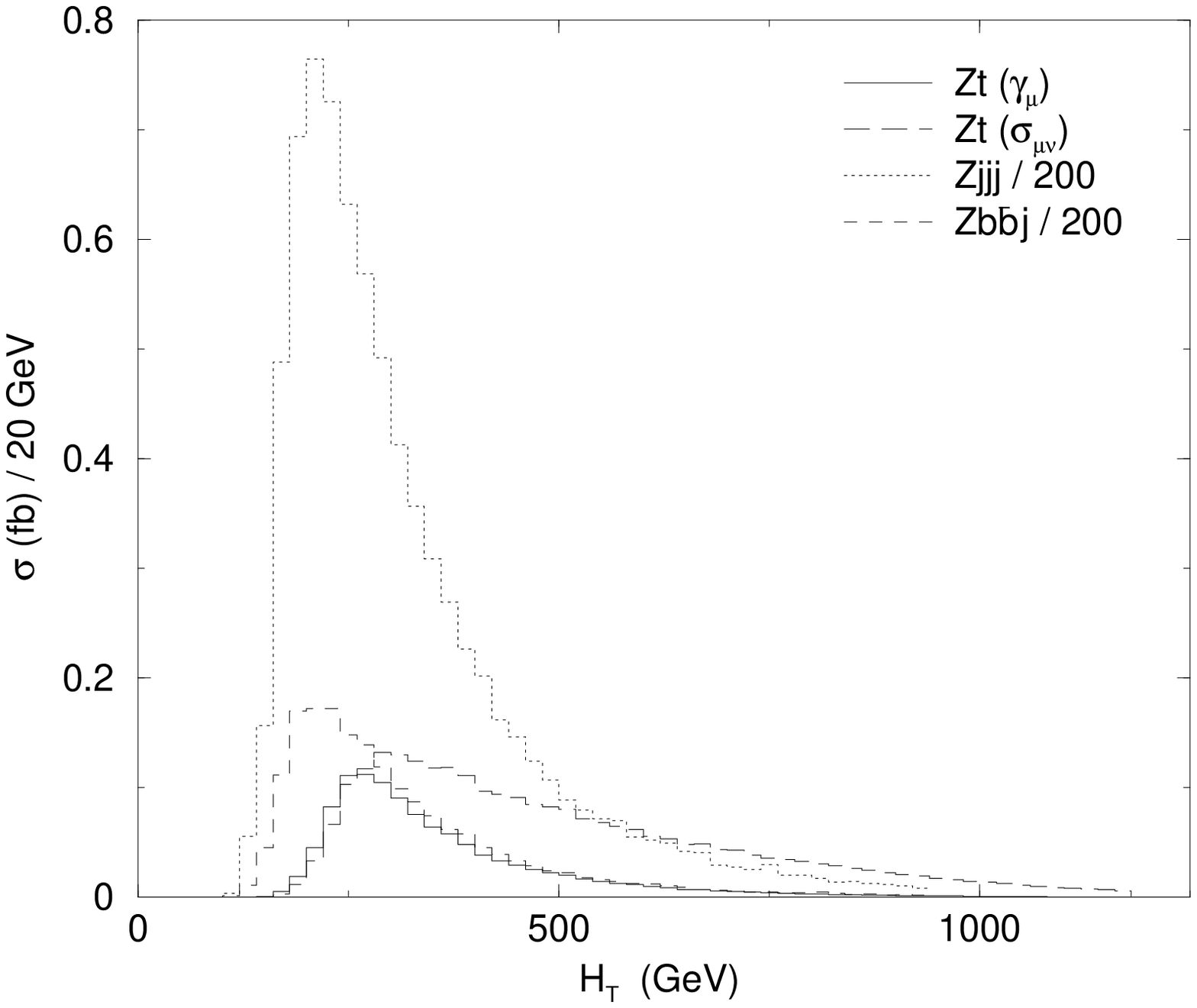,width=10cm}}
\caption{Total transverse energy $H_T$ distribution before kinematical cuts
for the $gu \to l^+ l^- jjb$ signal and backgrounds in LHC Run L. We use 
$X_{tu} = 0.02$, $\kappa_{tu}=0.02$ \label{fig:eejj-ht}}
\end{center}
\end{figure}

\begin{figure}[htb]
\begin{center}
\mbox{
\epsfig{file=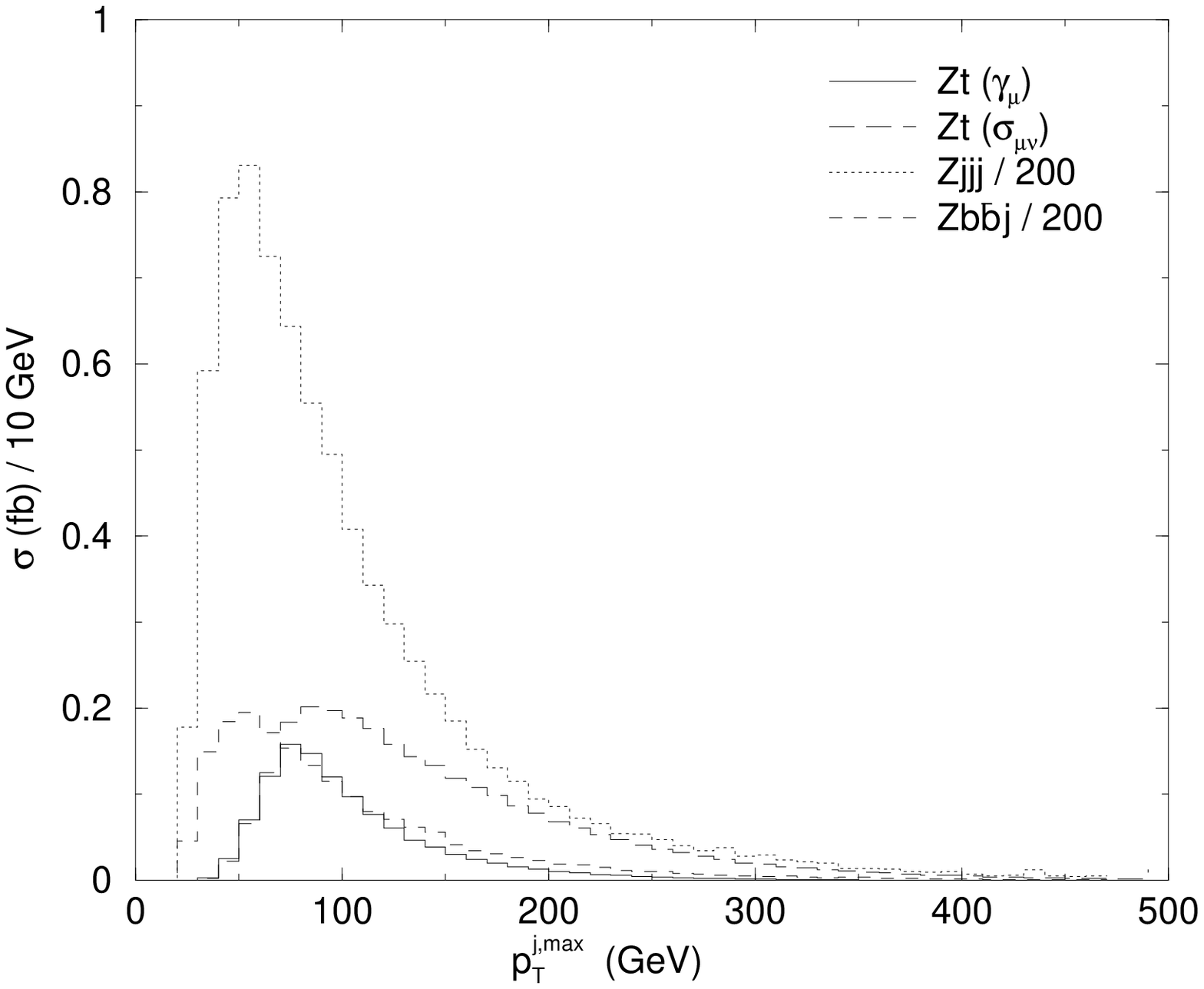,width=10cm}}
\caption{$p_T^{j,{\mathrm max}}$ distribution before kinematical cuts
for the $gu \to l^+ l^- jjb$ signal and backgrounds in LHC Run L. We use 
$X_{tu} = 0.02$, $\kappa_{tu}=0.02$ \label{fig:eejj-ptmax}}
\end{center}
\end{figure}

\begin{figure}[htb]
\begin{center}
\mbox{
\epsfig{file=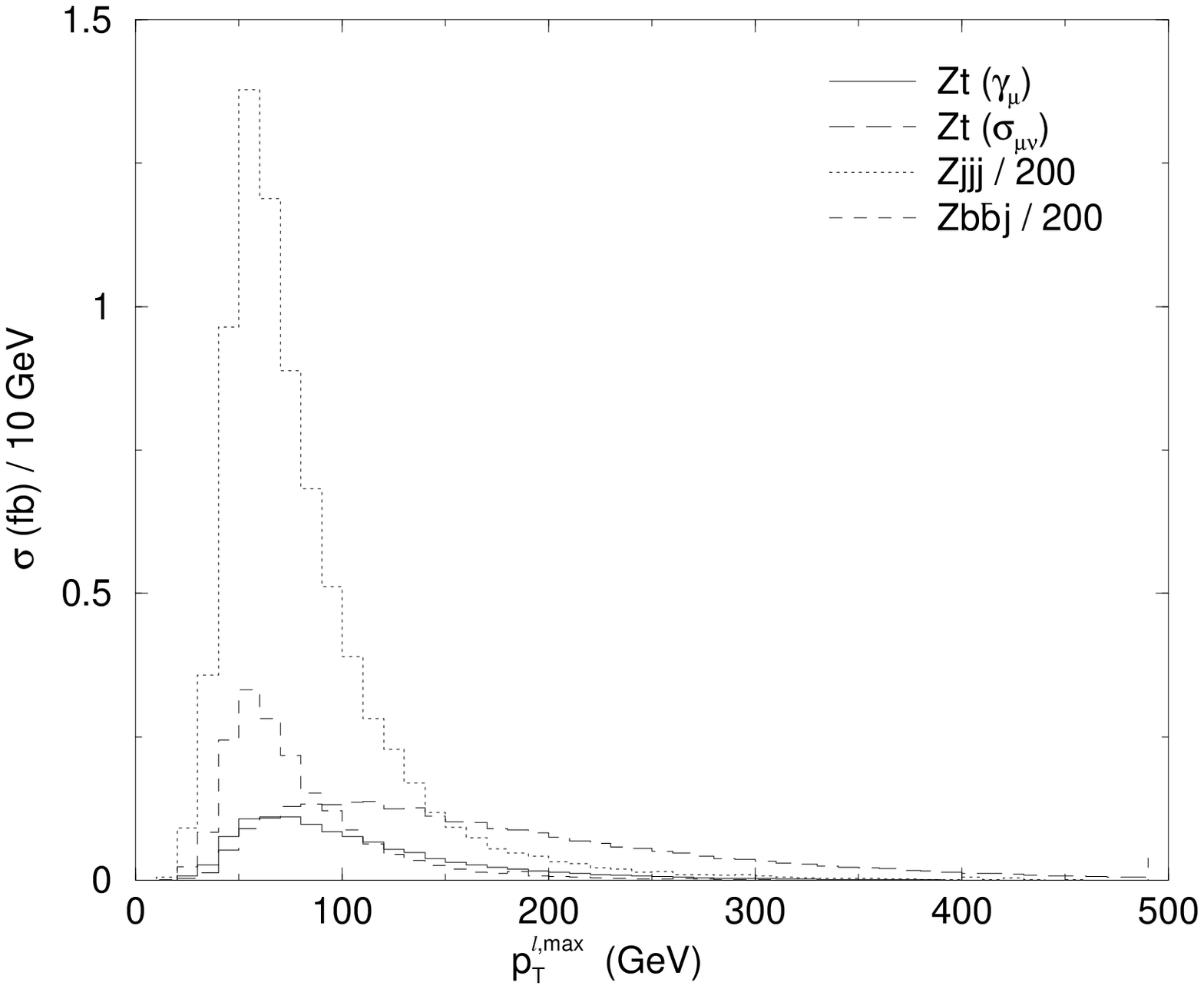,width=10cm}}
\caption{$p_T^{l,{\mathrm max}}$ distribution before kinematical cuts
for the $gu \to l^+ l^- jjb$ signal and backgrounds in LHC Run L. We use 
$X_{tu} = 0.02$, $\kappa_{tu}=0.02$ \label{fig:eejj-ptemax}}
\end{center}
\end{figure}

\begin{figure}[htb]
\begin{center}
\mbox{
\epsfig{file=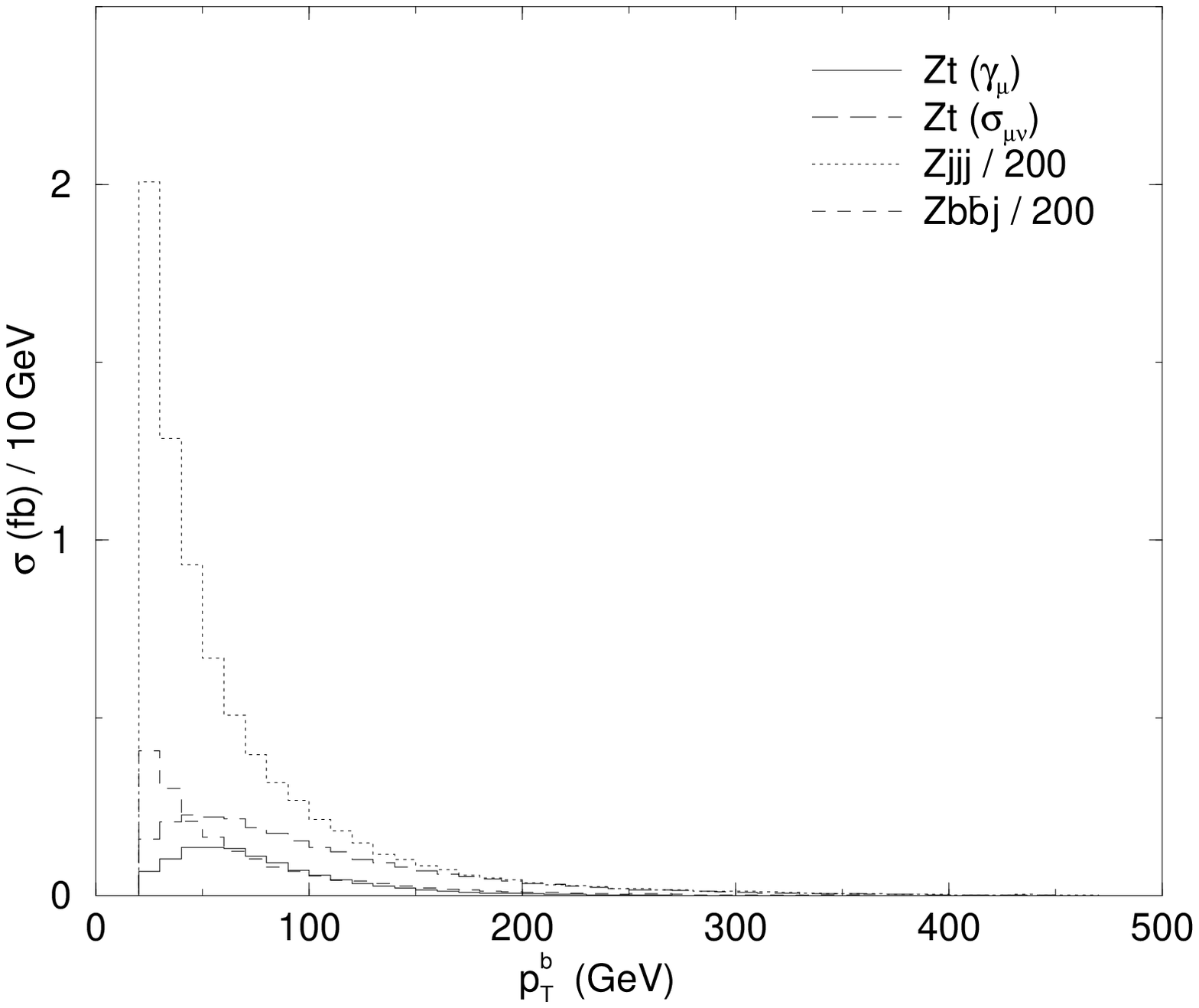,width=10cm}}
\caption{$p_T^b$ distribution before kinematical cuts
for the $gu \to l^+ l^- jjb$ signal and backgrounds in LHC Run L. We use 
$X_{tu} = 0.02$, $\kappa_{tu}=0.02$ \label{fig:eejj-ptb}}
\end{center}
\end{figure}

To reconstruct the signal we first perform $b$ tagging
with the corresponding probabilities of $0.6$ for $b$ jets and
$0.01$ for non $b$ jets, and require only one $b$
tag. This reduces the signal by a factor of 0.6, the largest background $Zjjj$
by 0.029 and the $Z b\bar b j$ background by 0.48. After tagging the $b$,
assumed to come from the top quark decay,
the two remaining jets are assigned to the $W$, and the $W$
reconstructed mass $\mwrec$ is defined by their invariant mass. In this case
the top
reconstructed mass $\mtrec$ is simply the invariant mass of the three jets.
These two invariant masses are not independent, 
$(\mtrec)^2 = (\mwrec)^2 + 2 p^W \cdot p^b$, and the kinematical cuts are
less effective than for the previous signal.
For $l^+ l^- jjb$ we also perform cuts
on $H_T$ and on the
transverse momenta of the fastest jet $p_T^{j,{\mathrm max}}$,
the fastest lepton $p_T^{l,{\mathrm max}}$ and the $b$ quark $p_T^b$.
All these
distributions are plotted in Figs.~\ref{fig:eejj-mw}--\ref{fig:eejj-ptb} for
the $gu \to Zt$ signal at LHC Run L.

We observe that the backgrounds are very concentrated
at low $p_T$'s. This makes convenient to use two different sets of cuts 1 and 2
for the $\gamma_\mu$ and $\sigma_{\mu \nu}$ couplings, given in
Table~\ref{tab:eejj1}. These cuts, especially that on $H_T$, are very
efficient for reducing the enormous background as can be observed in
Table~\ref{tab:eejj2}. For the $\sigma_{\mu \nu}$ couplings in Run H, requiring
very large transverse energy reduces the background by more than four orders of
magnitude
while retaining $26\%$ of the signal.
As in the previous Subsection, the leptonic trigger has little effect in Run H.
In fact this is smaller than the statistical
fluctuation of the Monte Carlo.
If no signal is observed, we find from Table~\ref{tab:eejj2}
$X_{tu} \leq 0.036$, $X_{tc} \leq 0.076$, $\kappa_{tu} \leq 0.015$,
$\kappa_{tc} \leq 0.045$ after Run L and $X_{tu} \leq 0.020$,
$X_{tc} \leq 0.041$, $\kappa_{tu} \leq 0.0076$, $\kappa_{tc} \leq 0.025$
after Run H.

\begin{table}[htb]
\begin{center}
\begin{tabular}{ccccc}
Variable & \multicolumn{2}{c}{Run L} & \multicolumn{2}{c}{Run H} \\
& Set 1 & Set 2 & Set 1 & Set 2 \\
$\mwrec$ & 70--90 & 70--90 & 70--90 & 70--90 \\
$\mtrec$ & 160--190 & 160--190 & 160--190 & 160--190 \\
$H_T$ & $>200$ & $>450$ & $>200$ & $>600$ \\
$p_T^{j,{\mathrm max}}$ & $>50$ & $>50$ & $>50$ & $>50$ \\
$p_T^{l,{\mathrm max}}$ & $>30$ & $>30$ & $>30$ & $>30$ \\
$p_T^b$ & & & & $>30$
\end{tabular}
\caption{Kinematical cuts for the $l^+ l^- jjb$ decay channel. The masses,
energies and momenta are in GeV. \label{tab:eejj1}}
\end{center}
\end{table}

\begin{table}[htb]
\begin{center}
\begin{tabular}{ccccccc}
 & \multicolumn{3}{c}{Run L} & \multicolumn{3}{c}{Run H} \\
& before & Set 1 & Set 2 & before & Set 1 & Set 2 \\[-0.4cm]
& cuts & cuts & cuts & cuts & cuts & cuts \\
$gu \to Zt (\gamma_\mu)$ & 11.3 & 9.9 & & 112 & 98.6 \\
$gc \to Zt\, (\gamma_\mu)$ & 2.6 & 2.2 & & 25.8 & 22.2 \\
$gu \to Zt\, (\sigma_{\mu \nu})$ & 26.5 & & 12.1 & 265 & & 68.6 \\
$gc \to Zt\, (\sigma_{\mu \nu})$ & 4.6 & & 1.4 & 46.3 & & 6.3 \\
$Zjjj$ & 15600 & 192 & 6.6 & 156000 & 1920 & 13.9 \\
$Zb \bar b j$ & 3660 & 42.5 & 1.5 & 35900 & 425 & 3.3 \\
$ZWj$ & 31.0 & 3.7 & 0.3 & 309 & 37 & 0.6
\end{tabular} 
\caption {Number of $l^+ l^- jjb$ events before and after
the kinematical cuts in Table~\ref{tab:eejj1} for the $Zt$ signal and
backgrounds. We use $X_{tq}=0.02$ and $\kappa_{tq}=0.02$.  \label{tab:eejj2}}
\end{center}
\end{table}

\subsection{$b \bar b l \nu b$ signal}

This decay mode has a larger branching ratio than $l^+ l^- l \nu b$,
but the need to tag two additional $b$'s and the trigger cuts reduce this
advantage. On the other hand the modes with few leptons have in general huge
backgrounds at LHC, which can be reduced mainly by $b$ tagging. If we
tag only one $b$ quark we are considering the signal $jj l\nu b$, which has
a branching ratio five times larger and gives nontrivial constraints at
Tevatron. However, at LHC energies the process $gb \to Wt \to jj l \nu b$ has a
large cross section and can mimic
the signal if the two non-$b$ jets have an
invariant mass consistent with the $Z$ mass. The 
$t \bar t \to WbW \bar b \to jjbl\nu \bar b$ cross section with one jet
missed is
even larger.
Requiring three $b$ tags both backgrounds become manageable. Other sizeable
background is the small $ZWj$ production. $Wb \bar b j$ production would be very
large if we would only require two $b$ jets. Requiring three $b$ tags reduces
this background to a moderate number of events due to the $b$ misidentification
factor of $0.01$. Finally $Wjjj$ becomes unimportant because it is suppressed by
a factor of $10^{-6}$ accounting for the three $b$ mistags.

In this channel we have to identify the two $b$ quarks 
resulting from $Z \to b \bar b$.
There are three pairs of tagged $b$ jets, and we choose the one with
invariant mass $\mzrec$ closest to $M_Z$. The other $b$ is assigned to the top
quark, and then the reconstructed top mass $\mtrec$ is calculated as for
$l^+ l^- l\nu b$. Other interesting variables are $H_T$ and $p_T^Z$.

A convenient set of cuts to improve the signal to background ratio is given in
Table~\ref{tab:bbb1}. In Table~\ref{tab:bbb2} we gather the number of
$b \bar b l \nu b$ events for the signal and backgrounds before and after these
kinematical cuts. We have used for the signal $X_{tq}=0.02$, $\kappa_{tq}=0.02$.
Note that although we have generate the $ZWj$ background with the $Z$ on-shell
and hence its $\mzrec$ distribution is sharply peaked around $M_Z$, this has
little effect because this background is small. From Table~\ref{tab:bbb2}
we obtain, if no signal is observed,
$X_{tu} \leq 0.056$, $X_{tc} \leq 0.12$, $\kappa_{tu} \leq 0.035$,
$\kappa_{tc} \leq 0.086$ after Run L and $X_{tu} \leq 0.035$,
$X_{tc} \leq 0.083$, $\kappa_{tu} \leq 0.019$, $\kappa_{tc} \leq 0.050$
after Run H.

\begin{table}[htb]
\begin{center}
\begin{tabular}{ccc}
Variable & Run L & Run H \\
$\mzrec$ & 80--100 & 80--100 \\
$\mtrec$ & 160--190 & 160--190 \\
$H_T$ & $>240$ & $>300$ \\
$p_T^Z$ & & $>50$ \\
\end{tabular}
\caption{Kinematical cuts for the $b \bar b l \nu b$ decay channel. The masses,
energies and momentum are in GeV. \label{tab:bbb1}}
\end{center}
\end{table}

\begin{table}[htb]
\begin{center}
\begin{tabular}{ccccc}
 & \multicolumn{2}{c}{Run L} & \multicolumn{2}{c}{Run H} \\
& before & after & before & after \\[-0.4cm]
& cuts & cuts & cuts & cuts \\
$gu \to Zt (\gamma_\mu)$ & 3.5 & 2.4 & 28.8 & 13.3 \\
$gc \to Zt\, (\gamma_\mu)$ & 0.8 & 0.5 & 6.3 & 2.4 \\
$gu \to Zt\, (\sigma_{\mu \nu})$ & 8.0 & 5.9 & 71.4 & 46.7 \\
$gc \to Zt\, (\sigma_{\mu \nu})$ & 1.4 & 1.0 & 11.9 & 6.7 \\
$t \bar t$ & 1790 & 62.8 & 14600 & 345 \\
$Wt$ & 21.0 & 8.3 & 161 & 36.2 \\
$ZWj$ & 11.1 & 2.4 & 93.1 & 8.3 \\
$Wb \bar b j$ & 115 & 1.5 & 924 & 4.3
\end{tabular} 
\caption {Number of $b \bar b l \nu b$ events before and after
the kinematical cuts in Table~\ref{tab:bbb1} for the $Zt$ signal and
backgrounds. We use $X_{tq}=0.02$ and $\kappa_{tq}=0.02$.  \label{tab:bbb2}}
\end{center}
\end{table}

\subsection{$\nu \bar \nu jjb$ signal}

Let us conclude this Section with a short discussion of the $\nu \bar \nu jjb$
channel. At Tevatron this is the most interesting mode due to its relatively
large branching
ratio, $13.6\%$, its moderate background and the relatively low luminosity of
the collider.
However, its $Wt$ and $t \bar t$ backgrounds
grow very quickly with energy (see Table~\ref{tab:tevlhc}) and make this
channel uninteresting at LHC.
We will then focus for comparison on the $Ztu$ couplings only.
The results for $Ztc$ are insignificant.

\begin{table}[htb]
\begin{center}
\begin{tabular}{cccc}
& Tevatron Run II & LHC Run L & Ratio \\
$gu \to Zt (\gamma_\mu)$ & 0.148 & 46 & 1:310 \\
$gu \to Zt\, (\sigma_{\mu \nu})$ & 0.173 & 112 & 1:640 \\
$Zjjj$ & 199 & 23200 & 1:120 \\
$Zb \bar b j$ & 74.1 & 4590 & 1:60 \\
$Wt$ & 3.5 & 18500 & 1:5300 \\
$t \bar t$ & 10.6 & 54800 & 1:5200
\end{tabular} 
\caption {Comparison between the number of $\nu \bar \nu jjb$ events without
kinematical cuts for the $Zt$ signal and
backgrounds at Tevatron and LHC.
We use $X_{tu}=0.02$ and $\kappa_{tu}=0.02$.  \label{tab:tevlhc}}
\end{center}
\end{table}

We reconstruct the signal as in the $l^+ l^- jjb$ case,
but with the charged leptons replaced by missing energy. We use for
both Runs the kinematical cuts in Table~\ref{tab:nunu1}. These are less
restrictive than for the $l^+ l^- jjb$ mode because the $Wt$ and $t \bar t$
backgrounds after missing the charged lepton and the $b$ quark
mimic the signal and are irreducible. The $Zjjj$ and $Z b \bar b j$
backgrounds are reduced by a factor of 80 (see Table~\ref{tab:nunu2}). The
bounds obtained 
are $X_{tu} \leq 0.068$, $\kappa_{tu} \leq 0.044$ after Run L
and $X_{tu} \leq 0.042$, $\kappa_{tu} \leq 0.021$ after Run H.

\begin{table}[htb]
\begin{center}
\begin{tabular}{ccccc}
Variable & Runs L and H \\
$\mtrec$ & 160--190 \\
$\mwrec$ & 70--90 \\
$H_T$ & $>180$ \\
\end{tabular}
\caption{Kinematical cuts for the $\nu \bar \nu jjb$ decay channel. The masses
and energy are in GeV. \label{tab:nunu1}}
\end{center}
\end{table}

\begin{table}[htb]
\begin{center}
\begin{tabular}{ccccccc}
 & \multicolumn{2}{c}{Run L} & \multicolumn{2}{c}{Run H} \\
& before & after & before & after \\[-0.4cm]
& cuts & cuts & cuts & cuts  \\
$gu \to Zt (\gamma_\mu)$ & 46.0 & 41.2 & 201 & 178 \\
$gu \to Zt\, (\sigma_{\mu \nu})$ & 112 & 101 & 759 & 681 \\
$t \bar t$ &  54800 & 46200 & 143500 & 123400 \\
$Wt$ & 18500 & 16100 & 41300 & 37300 \\
$Zjjj$ & 23200 & 279 & 76400 & 657 \\
$Zb \bar b j$ & 4590 & 67 & 14800 & 125 
\end{tabular} 
\caption {Number of $\nu \bar \nu jjb$ events before and after
the kinematical cuts in Table~\ref{tab:nunu1} for the $Zt$ signal and
backgrounds. We use $X_{tu}=0.02$ and $\kappa_{tu}=0.02$.  \label{tab:nunu2}}
\end{center}
\end{table}

\section{$\gamma t$ production}

In contrast with the $Zt$ case, $\gamma t$ production is not reduced by
branching fractions because the photon is stable. Moreover, since the photon is
massless, it tends to be produced with larger $p_T$'s. This effect is further
enhanced by the $q^\nu$ factor in the anomalous $\sigma_{\mu \nu}$ coupling.
Thus the larger $\gamma t$ cross section, in particular for large momenta,
allows for a better separation of the signal from the background, and then for
more precise measurements than in the previous cases.

There are two channels depending on the $W$ decay mode, $\gamma l \nu b$ and
$\gamma jjb$. We analyze them in turn. These processes constrain not only the
$\gamma t q$ anomalous couplings but also the strong anomalous couplings $gtq$.
The $\gamma l \nu b$ signal gives also in this case the most precise limits.
We also discuss them in detail.

\subsection{$\gamma l \nu b$ signal}

\begin{figure}[htb]
\begin{center}
\mbox{
\epsfig{file=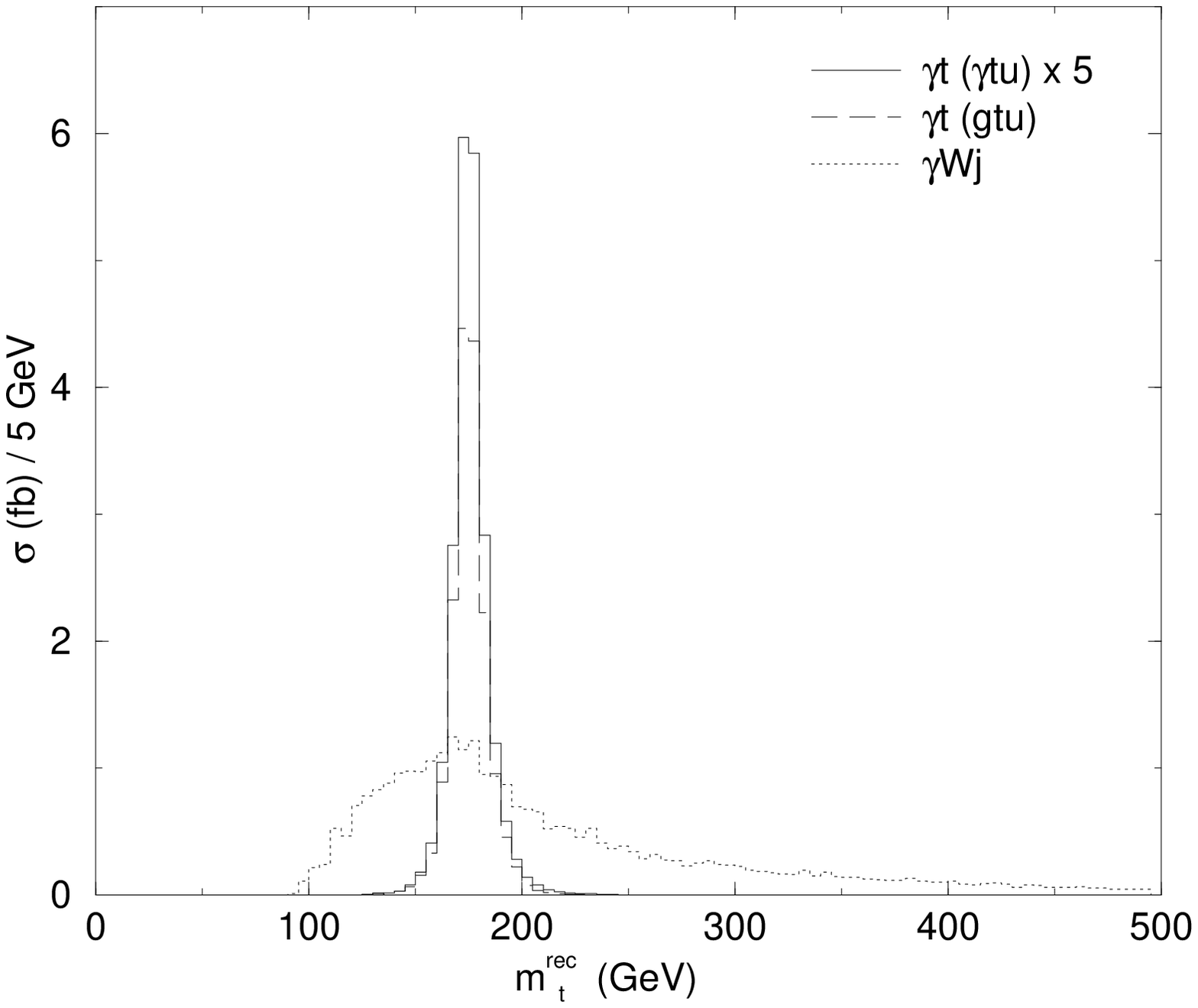,width=10cm}}
\caption{Reconstructed top mass $\mtrec$ distribution before kinematical cuts
for the $gu \to \gamma l \nu b$ signal and background in LHC Run L. We use 
$\lambda_{tu} = 0.01$, $\zeta_{tu}=0.01$ \label{fig:enub-mt}}
\end{center}
\end{figure}

\begin{figure}[htb]
\begin{center}
\mbox{
\epsfig{file=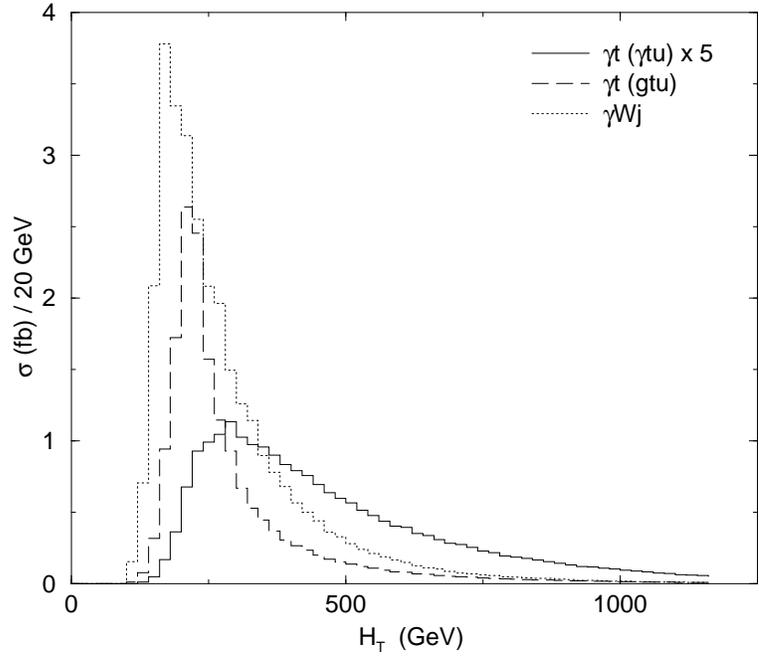,width=10cm}}
\caption{Total transverse energy $H_T$ distribution before kinematical cuts
for the $gu \to \gamma l \nu b$ signal and background in LHC Run L. We use 
$\lambda_{tu} = 0.01$, $\zeta_{tu}=0.01$ \label{fig:enub-ht}}
\end{center}
\end{figure}

\begin{figure}[htb]
\begin{center}
\mbox{
\epsfig{file=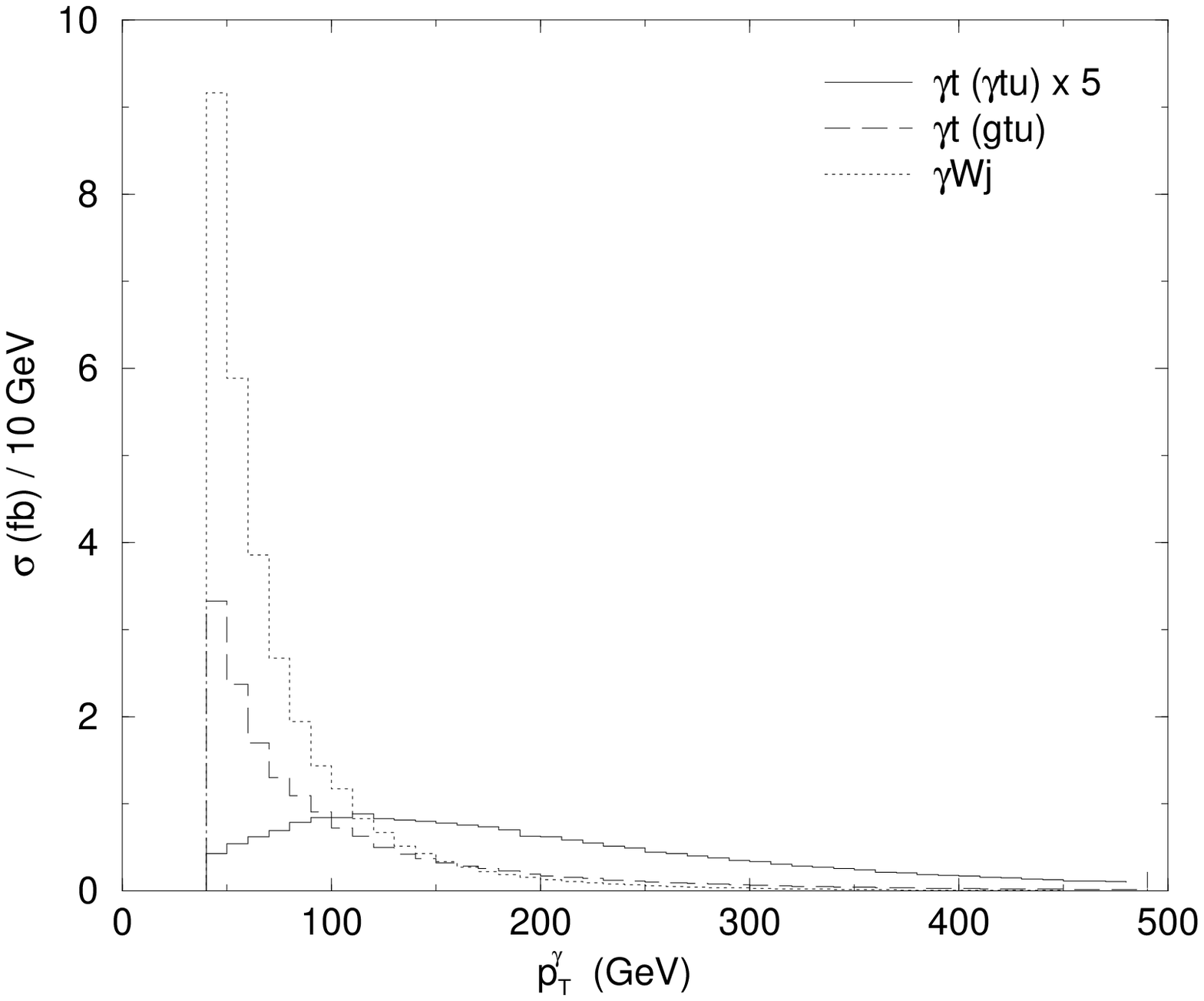,width=10cm}}
\caption{$p_T^\gamma$ distribution before kinematical cuts
for the $gu \to \gamma l \nu b$ signal and background in LHC Run L. We use 
$\lambda_{tu} = 0.01$, $\zeta_{tu}=0.01$ \label{fig:enub-ptgamma}}
\end{center}
\end{figure}

\begin{figure}[htb]
\begin{center}
\mbox{
\epsfig{file=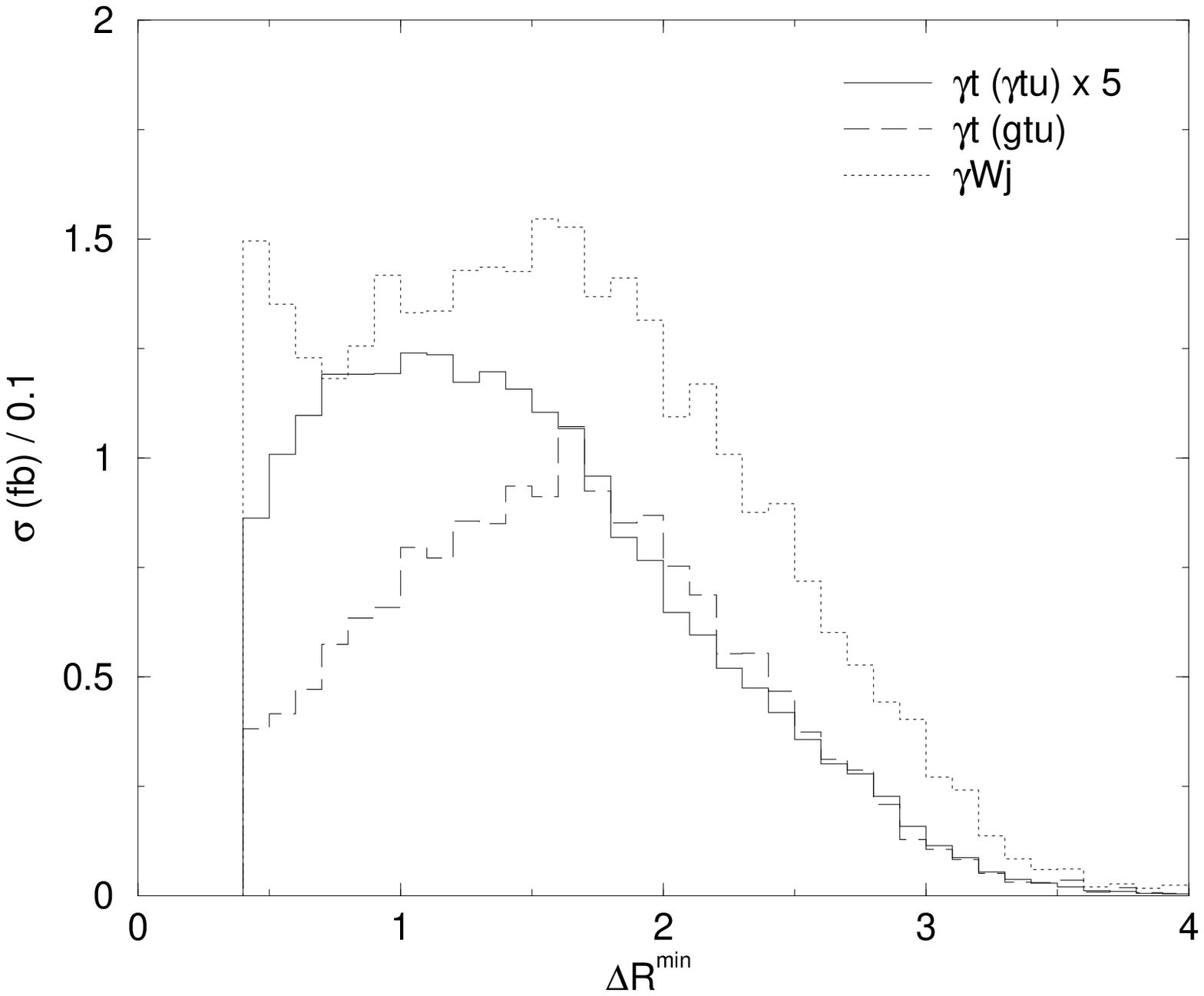,width=10cm}}
\caption{$\Delta R^{\mathrm min}$ distribution before kinematical cuts
for the $gu \to \gamma l \nu b$ signal and background in LHC Run L. We use 
$\lambda_{tu} = 0.01$, $\zeta_{tu}=0.01$ \label{fig:enub-drmin}}
\end{center}
\end{figure}

The leptonic $W$ decay gives a clean $\gamma l \nu b$ signal where the only
background is $\gamma Wj$ production with the jet misidentified as a $b$. This
case is then analogous to the $l^+ l^- l \nu b$ signal but
with the $l^+ l^-$ pair
replaced by the photon. The interesting
kinematical variables are $\mtrec$, defined as in the
$l^+ l^- l \nu b$ channel, $H_T$ and $p_T^\gamma$. The corresponding
distributions are plotted in
Figs.~\ref{fig:enub-mt}--\ref{fig:enub-ptgamma}. As emphasized above this signal
can also be produced via $gtq$ anomalous couplings. In this case the 
$\sigma_{\mu \nu}$ vertex couples the initial and not the final states in the
s-channel,
and the corresponding $\mtrec$, $H_T$ and $p_T^\gamma$ distributions in the same
Figures are different. Considering the minimum $\Delta R$ between the photon,
the charged lepton and the jet, $\Delta R^{\mathrm min}$, is also useful to
constrain these strong couplings. The signal for both processes and the
background distributions are shown
in Fig.~\ref{fig:enub-drmin}. It is then
convenient to apply different sets of cuts
for the electromagnetic and strong couplings. In Table~\ref{tab:enub1} we gather
both sets, 1 (2) for $\gamma tq$ ($gtq$). The total number of events before and
after these cuts are collected in Table~\ref{tab:enub2}, where we have used for
the signals $\lambda_{tq}=0.01$, $\zeta_{tq}=0.01$. 
If no signal is observed, we obtain 
$\lambda_{tu} \leq 0.0048$, $\lambda_{tc} \leq 0.013$,
$\zeta_{tu} \leq 0.0034$, $\zeta_{tc} \leq 0.0069$ after Run L and 
$\lambda_{tu} \leq 0.0021$, $\lambda_{tc} \leq 0.0065$,
$\zeta_{tu} \leq 0.0018$, $\zeta_{tc} \leq 0.0037$ after Run H.

\begin{table}[htb]
\begin{center}
\begin{tabular}{ccccc}
Variable & \multicolumn{2}{c}{Run L} & \multicolumn{2}{c}{Run H} \\
& Set 1 & Set 2 & Set 1 & Set 2 \\
$\mtrec$ & 160--190 & 160--190 & 160--190 & 160--190 \\
$H_T$ & $>300$ & $>200$ & $>300$ & $>200$ \\
$p_T^\gamma$ & $>100$ & & $>200$ \\
$\Delta R^{\mathrm min}$ & & $>0.6$ & & $>0.6$
\end{tabular}
\caption{Kinematical cuts for the $\gamma l \nu b$ decay channel. The masses,
energies and momenta are in GeV. \label{tab:enub1}}
\end{center}
\end{table}

\begin{table}[htb]
\begin{center}
\begin{tabular}{ccccccc}
 & \multicolumn{3}{c}{Run L} & \multicolumn{3}{c}{Run H} \\
& before & Set 1 & Set 2 & before & Set 1 & Set 2 \\[-0.4cm]
& cuts & cuts & cuts & cuts & cuts & cuts \\
$gu \to \gamma t\, (\gamma tu)$ & 43.3 & 29.3 & & 423 & 179 \\
$gc \to \gamma t\, (\gamma tc)$ & 7.5 & 4.2 & & 73.3 & 19.5 \\
$gu \to \gamma t\, (gtu)$ & 167 & & 118.5 & 1470 & & 1190 \\
$gc \to \gamma t\, (gtc)$ & 43.2 & & 29.1 & 368 & & 291 \\
$\gamma Wq_u$ & 140 & 2.5 & 16.8 & 1250 & 3.3 & 168\\
$\gamma Wq_d$ & 111 & 2.8 & 14.5 & 980 & 5.8 & 145\\
$\gamma Wg$ & 56.7 & 1.2 & 6.3 & 420 & 2.4 & 63
\end{tabular} 
\caption {Number of $\gamma l \nu b$ events before and after
the kinematical cuts in Table~\ref{tab:enub1} for the $Zt$ signal and
backgrounds. We use $\lambda_{tq}=0.01$ and $\zeta_{tq}=0.01$.
  \label{tab:enub2}}
\end{center}
\end{table}

\subsection{$\gamma jjb$ signal}

\begin{figure}[htb]
\begin{center}
\mbox{
\epsfig{file=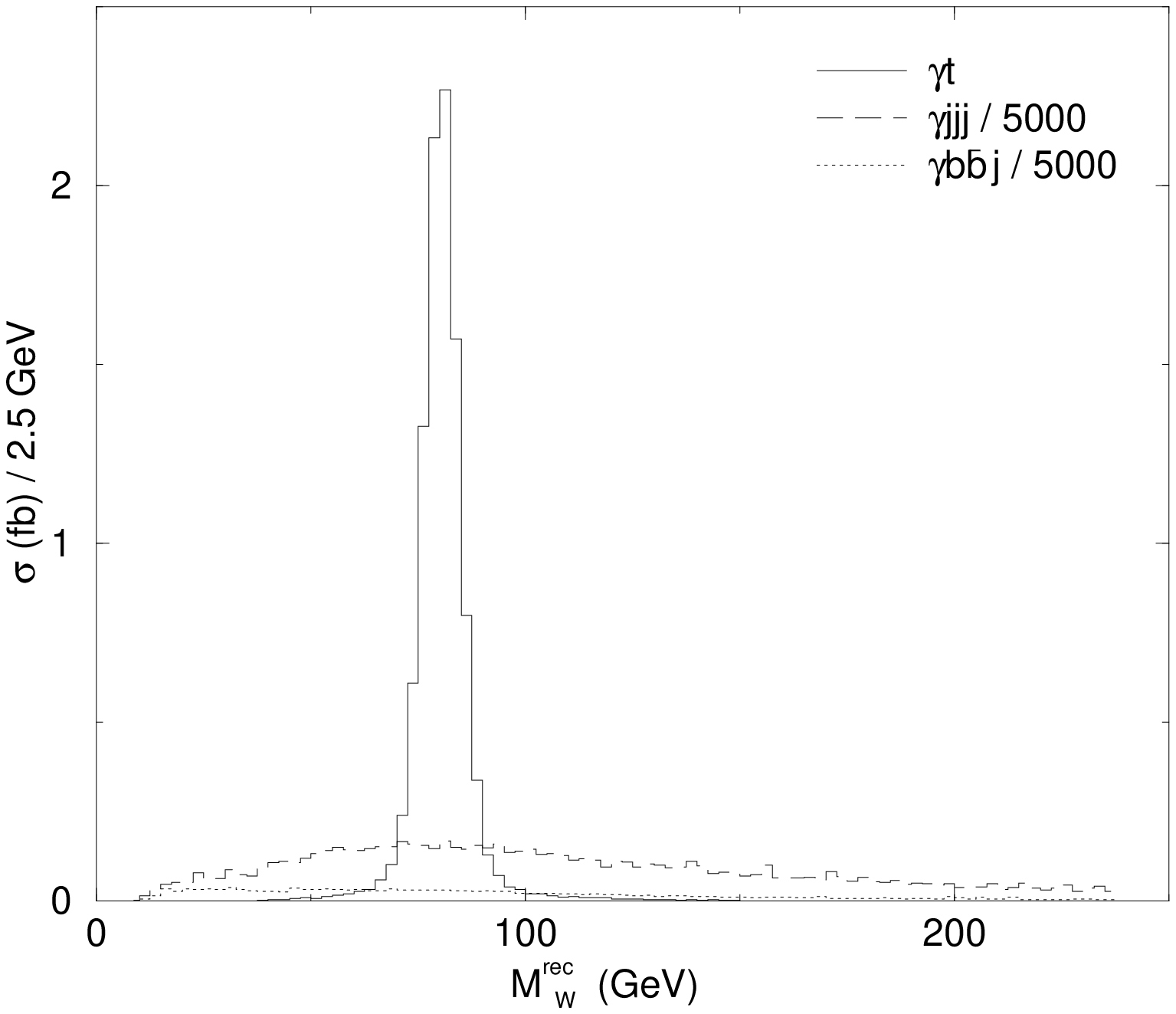,width=10cm}}
\caption{Reconstructed $W$ mass $\mwrec$ distribution before kinematical cuts
for the $gu \to \gamma jjb$ signal and backgrounds in LHC Run L. We use 
$\lambda_{tu} = 0.01$. \label{fig:jjb-mw}}
\end{center}
\end{figure}

\begin{figure}[htb]
\begin{center}
\mbox{
\epsfig{file=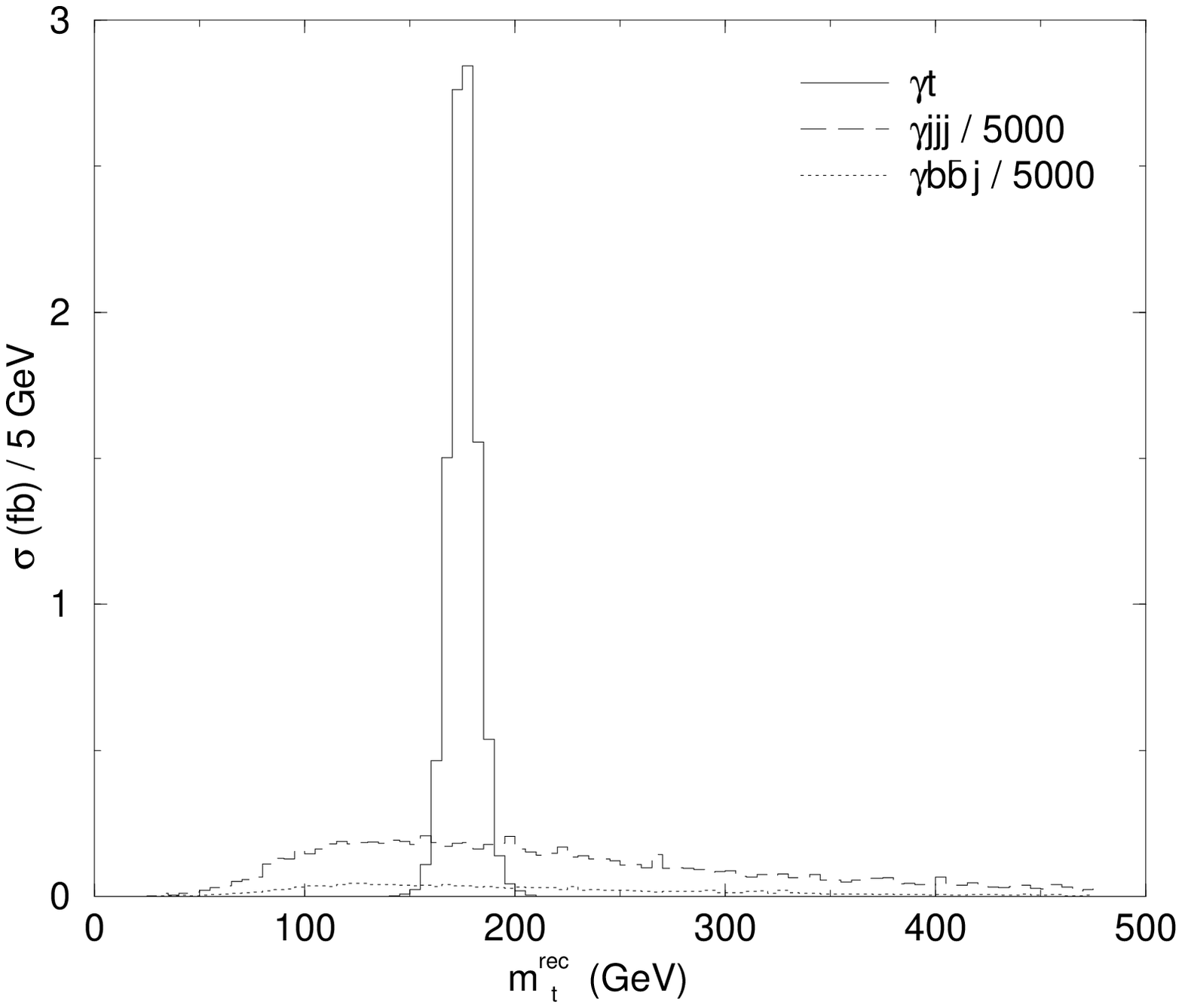,width=10cm}}
\caption{Reconstructed top mass $\mtrec$ distribution before kinematical cuts
for the $gu \to \gamma jjb$ signal and backgrounds in LHC Run L. We use 
$\lambda_{tu} = 0.01$. \label{fig:jjb-mt}}
\end{center}
\end{figure}

\begin{figure}[htb]
\begin{center}
\mbox{
\epsfig{file=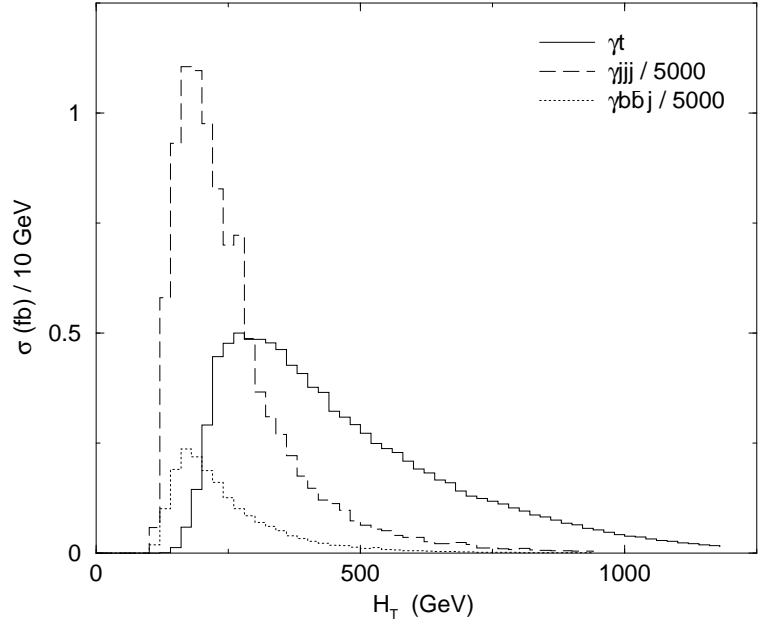,width=10cm}}
\caption{Total transverse energy $H_T$ distribution before kinematical cuts
for the $gu \to \gamma jjb$ signal and backgrounds in LHC Run L. We use 
$\lambda_{tu} = 0.01$. \label{fig:jjb-ht}}
\end{center}
\end{figure}

\begin{figure}[htb]
\begin{center}
\mbox{
\epsfig{file=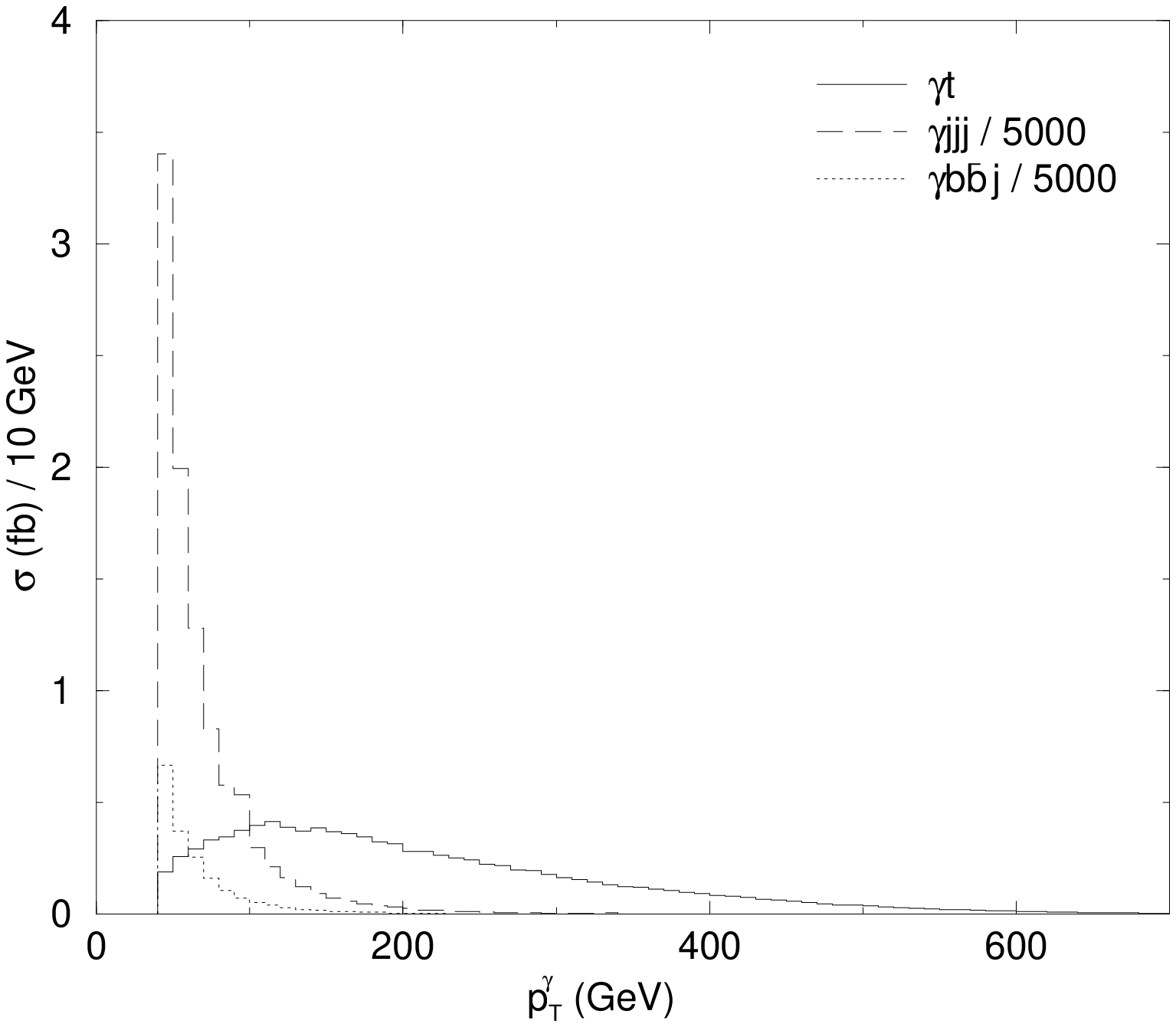,width=10cm}}
\caption{$p_T^\gamma$ distribution before kinematical cuts
for the $gu \to \gamma jjb$ signal and backgrounds in LHC Run L. We use 
$\lambda_{tu} = 0.01$. \label{fig:jjb-ptgamma}}
\end{center}
\end{figure}

\begin{figure}[htb]
\begin{center}
\mbox{
\epsfig{file=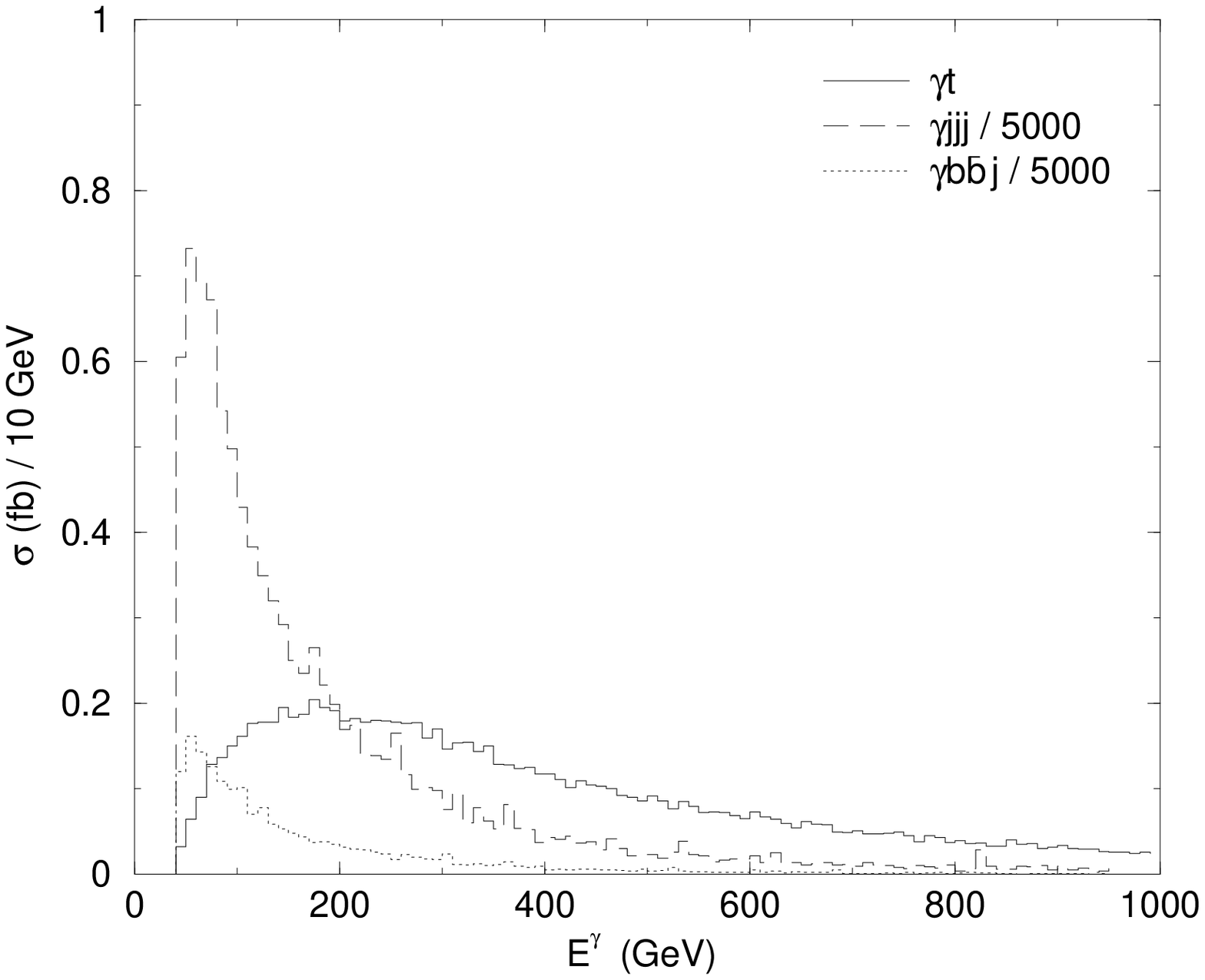,width=10cm}}
\caption{$E^\gamma$ distribution before kinematical cuts
for the $gu \to \gamma jjb$ signal and backgrounds in LHC Run L. We use 
$\lambda_{tu} = 0.01$. \label{fig:jjb-egamma}}
\end{center}
\end{figure}

\begin{figure}[htb]
\begin{center}
\mbox{
\epsfig{file=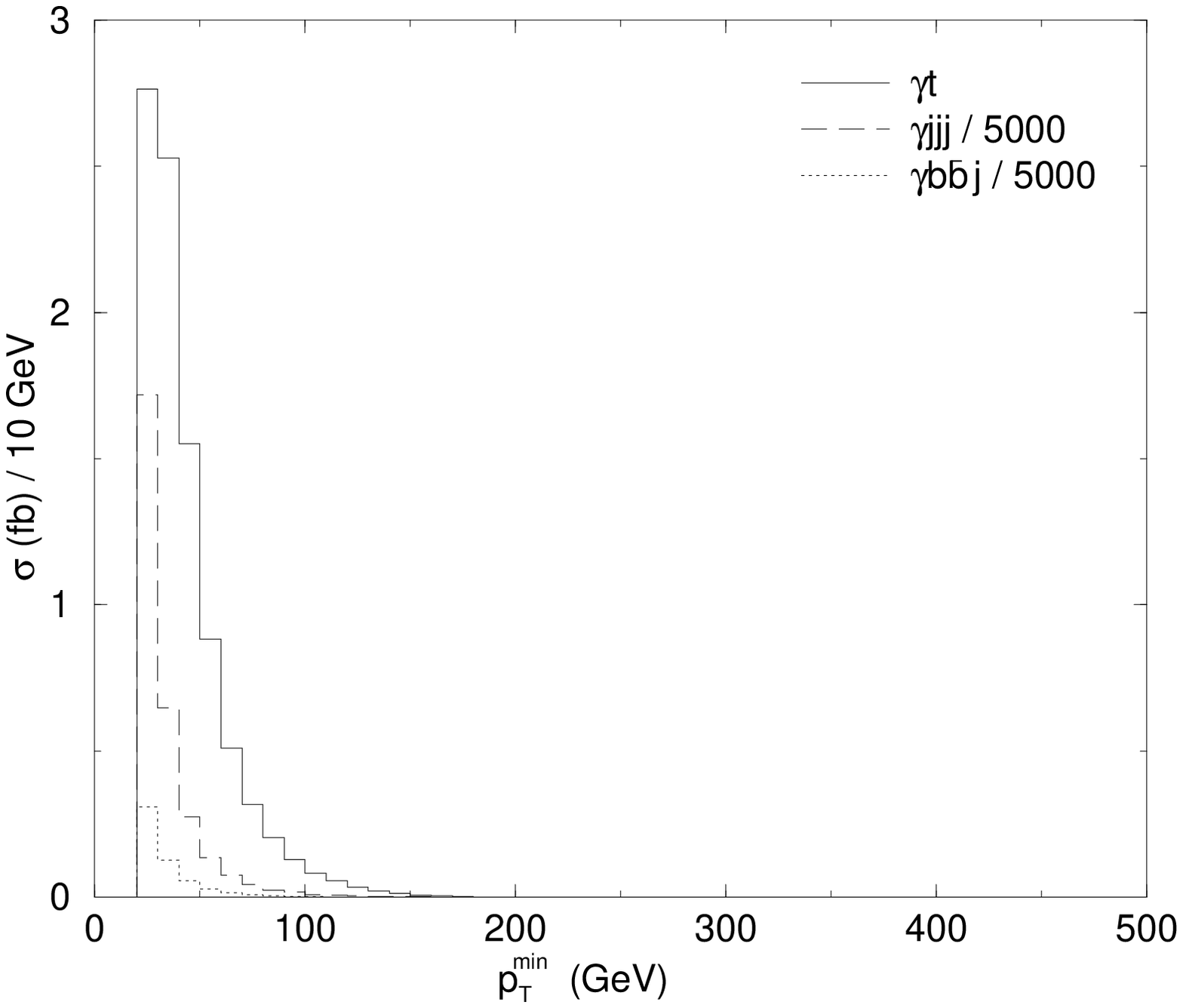,width=10cm}}
\caption{$p_T^{\mathrm min}$ distribution before kinematical cuts
for the $gu \to \gamma jjb$ signal and backgrounds in LHC Run L. We use 
$\lambda_{tu} = 0.01$. \label{fig:jjb-ptmax}}
\end{center}
\end{figure}

\begin{figure}[htb]
\begin{center}
\mbox{
\epsfig{file=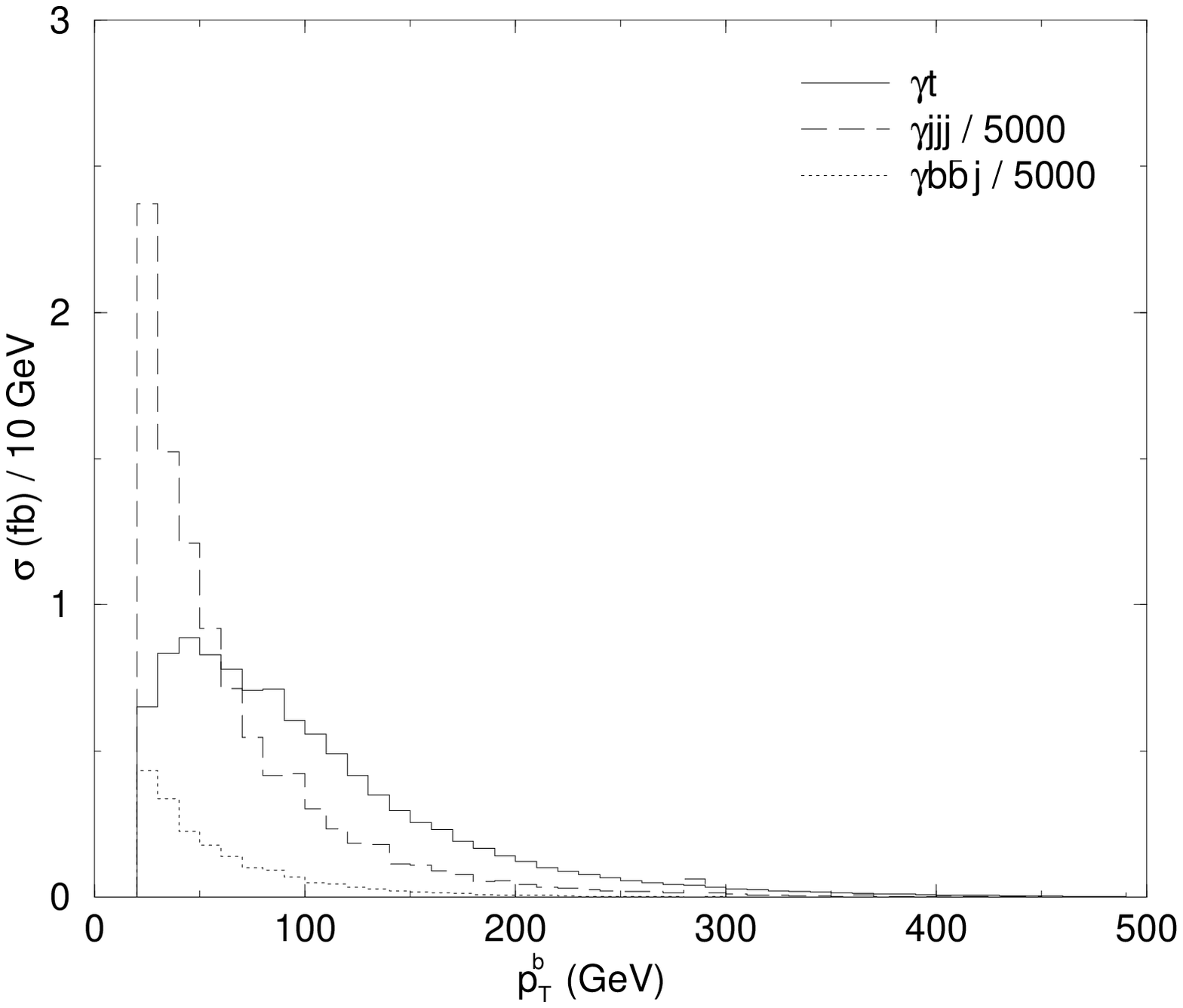,width=10cm}}
\caption{$p_T^b$ distribution before kinematical cuts
for the $gu \to \gamma jjb$ signal and backgrounds in LHC Run L. We use 
$\lambda_{tu} = 0.01$. \label{fig:jjb-ptb}}
\end{center}
\end{figure}

The $\gamma jjb$ mode is analogous to the $l^+ l^- jjb$ signal with the
lepton pair
replaced by the photon, and analogously to the $Zt$ case this channel is less
restrictive than the $\gamma l \nu b$ mode. Thus, although the hadronic $W$
branching ratio is larger than the leptonic one, the $\gamma jjb$ backgrounds
$\gamma jjj$ and $\gamma b \bar b j$ are
much larger than the $\gamma l \nu b$ ones.
In order to reduce them we exploit the $\sigma_{\mu \nu}$
behaviour of the signal and systematically require large momenta, {\em i. e.},
large $H_T$, $p_T^\gamma$, $p_T^b$, large photon energy $E^\gamma$ and minimum
jet transverse momentum $p_T^{\mathrm min}$, in addition to the usual
requirements on $\mwrec$ and $\mtrec$
(see Figs.~\ref{fig:jjb-mw}--\ref{fig:jjb-ptb}).
The signal distributions have the same
shape as for $\gamma l \nu b$, while the $\gamma jjj$ and $\gamma b \bar b j$
backgrounds are similar to $Zjjj$ and $Zb \bar b j$ but tend to be
more peaked at low $p_T$'s. This is due to the
masslessness of the photon. However, this effect
is reduced by the trigger requirement of
$p_T^\gamma \geq 40$.
A convenient set of kinematical cuts
is given in Table~\ref{tab:jjj1},
and the number of signal and background events before
and after applying
these cuts in Table~\ref{tab:jjj2}. We also take into account the
small $\gamma Wj$ background for comparison.
The effect of the kinematical cuts, especially at Run H, is impressive: the
background is reduced by more than $10^{-5}$ while retaining $5-10\%$ of the
signal. This allows to obtain competitive bounds at least on $\gamma t u$
couplings,
$\lambda_{tu} \leq  0.0059$, $\lambda_{tc} \leq 0.020$ after Run L and
$\lambda_{tu} \leq 0.0033$, $\lambda_{tc} \leq 0.012$ after Run H.

\begin{table}[htb]
\begin{center}
\begin{tabular}{ccc}
Variable & Run L & Run H \\
$\mwrec$ & 70--90 & 70--90 \\
$\mtrec$ & 160--190 & 160--190 \\
$H_T$ & $>540$ & $>700$ \\
$p_T^\gamma$ & $>230$ & $>300$ \\
$E^\gamma$ & $>300$ & \\
$p_T^{\mathrm min}$ & $>30$ & $>40$ \\
$p_T^b$ & $>50$ & $>60$
\end{tabular}
\caption{Kinematical cuts for the $\gamma jjb$ decay channel. The masses,
energies and momenta are in GeV. \label{tab:jjj1}}
\end{center}
\end{table}

\begin{table}[htb]
\begin{center}
\begin{tabular}{ccccc}
 & \multicolumn{2}{c}{Run L} & \multicolumn{2}{c}{Run H} \\
& before & after & before & after \\[-0.4cm]
& cuts & cuts & cuts & cuts \\
$gu \to Zt\, (\sigma_{\mu \nu})$ & 102 & 19.4 & 970 & 99.4 \\
$gc \to Zt\, (\sigma_{\mu \nu})$ & 18.5 & 1.7 & 170 & 7.9 \\
$\gamma jjj$ & 466000 & 6.9 & 2070000 & 15.3 \\
$\gamma b \bar b j$ & 92000 & 2.0 & 403000 & 5.2 \\
$\gamma Wj$ & 659 & 0.4 & 3430 & 1.0
\end{tabular} 
\caption {Number of $\gamma jjb$ events before and after
the kinematical cuts in Table~\ref{tab:jjj1} for the $Zt$ signal and
backgrounds. We use $\lambda_{tq}=0.01$.  \label{tab:jjj2}}
\end{center}
\end{table}

\section{Summary}

We have studied $Zt$ and $\gamma t$ production via top FCN couplings at LHC.
These processes manifest as 5 and 4 body final states, varying the statistical
significance of the different channels (see Table~\ref{tab:chan})
with the energy
and luminosity of the collider. Eventually at LHC the best limits on top FCN
couplings will be derived from $Zt \to l^+ l^- l \nu b$ and
$\gamma t \to \gamma l \nu b$. In Table~\ref{tab:best} we gather the
corresponding values for integrated luminosities of 10 fb$^{-1}$ (Run L) and
100 fb$^{-1}$ (Run H). In order to compare the reach of the different decay
modes, we collect in Table~\ref{tab:comp} the bounds on the $Ztu$
anomalous couplings
$X_{tu}$ and $\kappa_{tu}$ expected in Run H for the most significant signals.
At Tevatron with a luminosity of 109 pb$^{-1}$ (Run I) and 2 fb$^{-1}$ (Run II)
the most significant channels are $Zt \to \nu \bar \nu jjb$ and
$\gamma t \to \gamma l \nu b$ \cite{papiro2}. In both Runs the most sensitive
decay mode with the $Z$ boson decaying hadronically is $jj l \nu b$. This
variation of the relevance of the signals is mainly due to the small statistics
available at Tevatron and to the background increase at LHC, in part consequence
of the larger $b$ content of the proton. The statistics penalizes the channels
with few events when the backgrounds are negligible, whereas the large
backgrounds make uninteresting the less significant signals.

A few concluding remarks:
\begin{itemize}
\item If no signal is observed, $Zt$ and $\gamma t$ production at the LHC
will allow to
obtain independent bounds on the anomalous top couplings to up and charm
quarks.

\item Limits derived from top decays are less precise but comparable, especially
taking into account that they have been obtained using a conservative estimate.

\item The bounds on the anomalous $\sigma_{\mu \nu}$ couplings in
Table~\ref{tab:best} seem more stringent than those on $\gamma_\mu$ couplings
because they are normalized to $m_t$ which we take equal to 175 GeV in
Eq.~(\ref{ec:1}), whereas the energies
probed are significantly larger. If instead
we would have used $\Lambda=1$ TeV, the
constraints on $\sigma_{\mu \nu}$ couplings would have looked less stringent.

\item Strong anomalous couplings $gtq$ are more precisely
 constrained by single $t$ production
\cite{papiro17}.

\item All these bounds may be too optimistic at the end and a real simulation of
the experimental conditions without neglecting a priori other possible
small backgrounds and the uncertainties associated with the structure functions
is necessary to obtain better estimates.

\item The $Vt$ signal is understimated because $t \bar t$ production with a FCN
top decay into $Wb$ and an undetected light quark is not included.

\item Multilepton signals are expected from other processes,
for example they are characteristic in gauge
mediated supersymmetry breaking models \cite{papiro27}. However if their origin
is $Zt$ and $\gamma t$ production,
the fixed ratios between the different $Z$ and
$t$ decay modes will allow to establish their origin.

\item The best limits on top FCN couplings are expected to be obtained at
LHC because it will produce many more tops than other planned machines.
However, the clean
environment of $e^+ e^-$ colliders makes them complementary, particularly if
new physics is observed.

\item In any case, $b$ tagging plays an essential r\^{o}le in tracing tops and
reducing backgrounds.

\item The large number of tops to be produced at future colliders,
the present lack of precise
knowledge of the top properties and the widely spreaded idea that new physics
must first manifest in the heaviest family make top physics
particularly important.

\item In a general effective Lagrangian approach with only the SM light
degrees of freedom the lowest order top FCN couplings are all dimension 6
\cite{papiro10,papiro3}.
After
electroweak symmetry breaking they generate the dimension 4 and 5 couplings we
have considered, which are the lowest dimension top FCN vertices with only one
gauge boson. Thus, testing these couplings one expects to probe a large class of
SM extensions.
\end{itemize}

\begin{table}
\begin{center}
\begin{tabular}{llcccccccc}
Signal & Run & $X_{tu}$ & $X_{tc}$ & $\kappa_{tu}$ & $\kappa_{tc}$ &
$\lambda_{tu}$ & $\lambda_{tc}$ & $\zeta_{tu}$ & $\zeta_{tc}$ \\
$l^+ l^- l \nu b$ & Run L & 0.022 & 0.045 & 0.014 & 0.034 &
--- & --- & 0.0069 & 0.017 \\
& Run H & 0.011 & 0.023 & 0.0063 & 0.016 &
--- & --- & 0.0030 & 0.0078 \\
$\gamma l \nu b$ & Run L & --- & --- & --- & --- &
0.0048 & 0.013 & 0.0034 & 0.0069 \\
& Run H & --- & --- & --- & --- &
0.0021 & 0.0065 & 0.0018 & 0.0037 \\
\end{tabular}
\caption{Most stringent bounds on the anomalous top couplings in
Eqs.~(\ref{ec:1}), (\ref{ec:2})
from single top production in
association with a $Z$ boson or a photon at the LHC.  \label{tab:best} }
\end{center}
\end{table}

\begin{table}
\begin{center}
\begin{tabular}{lcc}
Channel & $X_{tu}$  & $\kappa_{tu}$ \\
$l^+ l^- l \nu b$   & 0.011 & 0.0063 \\
$l^+ l^- jjb$       & 0.020 & 0.0076 \\
$ b \bar b l \nu b$ & 0.035 & 0.019 \\
$ \nu \bar \nu jjb$ & 0.042 & 0.021
\end{tabular}
\caption{Limits on the $Ztu$ anomalous couplings $X_{tu}$ and $\kappa_{tu}$
for the most significant decay channels at LHC Run H.}
\label{tab:comp}
\end{center}
\end{table}

\begin{ack}
We thank W. Giele for helping us with VECBOS and J. Fern\'{a}ndez
de Troc\'{o}niz
and I. Efthymiopoulos for discussions on Tevatron and LHC triggers. We have also
benefited from discussions with F. Cornet, M. Mangano and R. Miquel and from
previous collaboration with Ll. Ametller.
This work was partially supported by CICYT under contract AEN96--1672 and by the
Junta de Andaluc\'{\i}a, FQM101.
\end{ack}

\newpage

\newpage

\newpage

\newpage

\newpage

\newpage

\newpage

\newpage

\newpage

\newpage

\newpage

\end{document}